\documentclass[preprints,article,accept,pdftex,moreauthors]{mdpi}
\firstpage{1} 
\pubvolume{1}
\issuenum{1}
\articlenumber{0}
\pubyear{2022}
\copyrightyear{2022}
\datereceived{} 
\dateaccepted{} 
\datepublished{} 
\hreflink{https://doi.org/} % If needed use \linebreak
\Title{Diffusion and velocity correlations of the phase transitions in a system of macroscopic
  rolling spheres}

\TitleCitation{Title}

\Author{F. Vega Reyes $^{1,2}$\orcidA{}, A. Rodr\'iguez-Rivas $^{3}$\orcidB{},
  J. F. Gonz\'alez-Saavedra $^{1}$\orcidC{} and M. A. L\'opez-Casta\~no $^{1}$\orcidD{}}

\AuthorNames{Firstname Lastname, Firstname Lastname and Firstname Lastname}

\AuthorCitation{ Vega Reyes, F.; Rodr\'iguez-Rivas, A.; J. F. Gonz\'alez-Saavedra; L\'opez-Casta\~no,
  M. A.} 
\address{%
  $^{1}$ \quad Departamento de F\'isica, Universidad de Extremadura, 06071, Badajoz, Spain \\
  $^{2}$ \quad Instituto de Computaci\'on Cient\'ifica Avanzada (ICCAEx), Universidad de
  Extremadura, 06071, Badajoz, Spain \\
  $^{3}$ \quad Department of Physical, Chemical and Natural Systems, Pablo de Olavide University,
  41013, Sevilla, Spain\\ }

\corres{Correspondence: fvega@eaphysics.xyz (F. V. R.)}

\abstract{We study an air-fluidized granular monolayer, composed of plastic spheres
  which roll on a metallic grid. The air current is adjusted so that the spheres never loose contact
  with the grid, so that the dynamics may be regarded as pseudo two-dimensional (or two-dimensional,
  if the effects of sphere rolling are not taken into account). We find two surprising continuous
  transitions, both of them displaying two coexisting phases. Moreover, in all cases, we found the
  coexisting phases display strong energy non-equipartition. In the first transition, at weak
  fludization, a glassy phase coexists with a disordered fluid-like phase. In the second transition,
  a hexagonal crystal coexists with the fluid phase. We analyze, for these two-phase systems, the
  specific diffusive properties of each phase, as well as the velocity correlations. Surprisingly,
  we find a glass phase at very low packing fraction and for a wide range of granular
  temperatures. Both phases are characterized also by a strong anti-correlated velocities upon
  collision. Thus, the dynamics observed for this quasi two-dimensional system unveils phase
  transitions with peculiar properties, very different from the predicted behavior in well know
  theories for their equilibrium counterparts.}


\begin{document}

%%%%%%%%%%%%%%%%%%%%%%%%%%%%%%%%%%%%%%%%%%
%\setcounter{section}{-1} %% Remove this when starting to work on the template.
\section{Introduction}

The dynamics of granular matter has been an emerging field for several decades now
\cite{JNB96,dG99}. This is partly due to the many industrial and engineering applications that this
kind of materials has \cite{AT06}; and partly due the fact that granular set-ups can be used as
prototype non-equilibrium systems for experiments \cite{OU98}, and also, from a theoretical
viewpoint, they allow for the development of the theory of non-equilibrium statistical mechanics
\cite{G03}, fluid mechanics \cite{VU09} and materials science \cite{LAYKA14,GRLV2020}. Moreover,
advances on granular dynamics theory have clearly put in evidence that, both at mesoscopic and
macroscopic level, the dynamics of granular matter \cite{JNB96,G03,Mujica2016} can present analogous
phenomenology to that of molecular matter but is usually present in more complex ways. This is the
case, for instance, of phenomena such as stratification \cite{AT06}, diffusion
\cite{ZS91,Oger1996,OLDLD04}, segregation \cite{RSPS87,KHTPB03,JY02,HKGMO99}, mixing
\cite{Melby2005,AT06}, laminar flow \cite{VU09}, hydrodynamic instabilities \cite{EWMBL07,MKPM19},
convection \cite{XMD02,EWMBL07,PGPV17}, turbulence \cite{I12a,I03}, jamming \cite{LN98,Daniels2012},
memory effects \cite{Lasanta2017,KPZSN19} and phase transitions
\cite{PMEU04,Melby2005,RIS06,VU08,CMS12,NRTMS14,CMS15}, just to name a few. With respect to phase
transition, granular matter displays disordered states which can be for instance liquid-like, glassy
or hyperuniform states, and ordered structures like nematic phases, hexagonal or cubic crystals. It
also shows phases that are exclusive of two-dimensional (2D) systems, like the hexatic phase.

As an example of the higher complexity of granular dynamics, and just out of illustration, the set
of steady base flows that can be observed in a plane Fourier/Couette configuration (a fluid confined
within two infinite parallel walls) includes those that are present in molecular gases plus new
steady flows, that are specific of granular fluids \cite{VU09}. In particular, the Fourier
configuration (two static parallel walls) for a molecular gas yields steady flows with constant heat
flux; these constant heat flux states are however possible in a granular gas if the confining
parallel walls are moving (Couette configuration) \cite{VSG10}. Moreover, in addition to this
complex phenomenology that is also present in simpler forms in molecular matter, there are phenomena
in granular matter that do not have an analog in their equilibrium counterparts, like granular
nucleation \cite{RRSS18}, inelastic clustering \cite{GZ93}, collapse \cite{OU98} or velocity
correlations that, at low density, clearly violate the molecular chaos assumption \cite{PEU02}. So,
granular dynamics can be regarded, from the theoretical point of view, as an extension or
generalization of the dynamics of molecular matter \cite{G03,MS16}.

In this work, we focus instead in the phase transitions, and order/disorder phenomenology in a
monolayer of macroscopic spheres. Spheres are fluidized by turbulent air currents, in such a way
that they keep rolling over a horizontal plane at all times. In this way, the activated dynamics
remains as quasi-2D (or pseudo-2D, as preferred). On the other hand, and as it is well known, an
equilibrium fluid in two-dimensions crystallizes to a hexagonal phase via a continuous transition
that is mediated by a phase that is specific of two dimensions (the hexatic phase). This process is
well described by the KTHNY scenario (from their main authors K\"osterlitz, Thouless, Halperin,
Nelson, Young; see their independent works \cite{Kosterlitz1972,KT73,NH79,Y79}). The hexatic
phase appears exclusively in 2D and is characterized by having quasi-long-ranged orientational
correlations (with power-law slow decay) and short-ranged translational correlations (with
exponential decay). By contrast, the hexagonal crystal shows quasi-long-ranged translational
order. Thus, as the crystal melts, long translational order is lost, and correlations undergo a
complete transformation process towards complete disorder, which characterizes the liquid phase
\cite{S88}.

This liquid-hexatic-crystal scenario has been observed in non-equilibrium systems as well, although
with (eventually) important variations. For instance, the hexatic phase has been detected in a
monolayer of vertically vibrated macroscopic spheres \cite{OU05,KT15}. More specifically, in the
work by Olafsen \& Urbach \cite{OU05}, the phenomenology for the quasi-2D non-equilibrium system
appears to be rather similar to that described for equilibrium systems that are strictly
two-dimensional. To the point that virtually no difference with respect to the equilibrium theory
was detected. In effect, a first transition was found, where the crystal melts to the intermediate
hexatic phase described by the KTHNY scenario, by means of a process of unbinding of dislocation
pairs in the hexagonal lattice. A second transition consecutively occurs, where the hexatic phase
decays to a liquid-like phase with complete disorder, through a process of gradual unbinding of
disclinations. In this way, the crystal melting process occurs as a double continuous transition,
without a coexistence with the liquid phase ever taking place, unlike in three-dimensional
matter. In the work by Olafsen \& Urbach \cite{OU05}, however, experiments were performed only with
stainless steel spheres, which are nearly elastic (see the work by Louge and collaborators
\cite{FLCA94}, accompanied with the comprehensive data table \cite{FLCA_data94}, where the
coefficients of restitution for steel and other metals are given). In fact, the corresponding
velocity distribution shows little to no deviations off the Maxwellian distribution
\cite{OU99,OU01}, meaning that the system is not far from equilibrium \cite{Brey1998,BC01}. In this
way, one can say that for the phase behavior might be expected not to differ much from that of a
truly equilibrium system. In this sense, further experiments, performed by Komatsu \& Tanaka
\cite{KT15}, with the same monolayer configuration found an intriguing disappearance of the KTHNY
scenario for rubber spheres, which are more inelastic than steel spheres \cite{FLCA_data94}. They
observed an abrupt change from continuous to discontinuous melting. In fact, the melting transition
for very inelastic spheres was found exhibit phase coexistence between the crystal and the liquid,
without the hexatic phase ever showing up in the process; i.e., the KTHNY scenario disappears
completely at high inelasticity.  (In fact a previous result for brass spheres, also more inelastic
than stainless steel \cite{FLCA_data94}, in a two-layer system, already proved that increasing the
degree of inelasticity can alter importantly the phase behavior of the granular layer \cite{VU08}.)

Furthermore, the 2D phase behavior in active matter seems to be even more complex than in granular
matter. In fact, a complex m\'elange between the KTHNY scenario with the motility-induced phase
separation (that is characteristic of active matter \cite{BLLRVV16}) can be observed for wide ranges
of particle density and at strong particle activity \cite{DLSCGP18}. In summary, according to strong
experimental and computational evidence, the KTHNY scenario is only one of the possible realizations
of the phase behavior in 2D or nearly 2D non-equilibrium systems, and frequently appears mixed with
other transitions. Moreover, these alternative scenarios in equilibrium systems seem to be depend
upon the type of interaction between the particulate system constituents \cite{S88}. In this sense,
the KTHNY scenario is not the only possible phase behavior in two-dimensional systems was in fact
also predicted for equilibrium systems as well. In effect, first order transitions in 2D equilibrium
systems seem to be viable, according to theoretical analysis and Monte Carlo simulations, if
dislocation and disclination unbinding are concurrent \cite{S88}. Moreover, this concurrence appears
to depend upon the features of the interactions between the particles of the system. Therefore,
particle interactions seem to play an important role in the phase behavior in 2D.

The present work, which deals with a system of air-fluidized ping-pong balls, is motivated by the
previous discussion on the phase behavior in 2D particulate systems. In this sense, previous
experimental observation puts in evidence the existence of long-ranged repulsive interactions
between the air-fluidized particles \cite{OAD05}. This interaction, whose origin lies in the usual
hydrodynamic interactions due to the presence of the interstitial fluid \cite{B74} (air in this
case), is absent in the case of the vibrated system. Moreover, this repulsion does not prevent
particle direct encounters (collisions) \cite{Ojha2005,LGRAYV21} . Thus, due to the combined action
of inelastic collisions plus repulsive potential between particles, a different phase behavior could
be expected, according to the theoretical discussion above for their equilibrium analogs. Analysis
of eventual departures from the KTHNY scenario when long-ranged hydrodynamic interactions would be
relevant for a more complete understanding of the phase behavior in two-dimensional
systems. Although this experimental configuration has been studied in a number of previous works
\cite{OAD05,Abate2005,LGRV20,LGRAYV21,KMN21}, they have focused on different features of the complex
dynamics that this type of system exhibits, and no detailed analysis has been carried out on the
phase behavior, except for some previous work where a generic description has been provided
\cite{LGRAYV21,KMN21}.

For this reason, we focus in this work on the specific features of the observed phases and the
transitions between them. As we will see, several discontinuous phase transitions can clearly be
detected as air upflow intensity is increased, giving rise to states with a either a single phase or
two coexisting phases. We will study the granular temperature field (here defined as
$T=(1/2m\langle v^2\rangle)$, where $\langle v^2\rangle$ the square of particle velocity spatially
averaged over all steady states), as well as the structure (particle density, pair correlation
function) and dynamical properties (diffusion and velocity autocorrelations) of each of the observed
phases. The combined analysis of these magnitudes allows us for identifying the the following
phases: arrest phase, glass, liquid and hexagonal crystal. In particular, we show that the glass and
the crystal phases are clearly subdiffusive. Surprisingly, the liquid phase can display either
normal diffusion or weakly subdiffusive (or superdiffusive) behavior. As we will see, these
transitions in the diffusive behavior of the system occur in a discontinuous way. Furthermore, in
the glassy phase, particle velocities are strongly anticorrelated at early times, whereas the
crystal anticorrelations are weak. We also found strong energy non-equipartition in all cases of two
coexisting phases.

The paper is structured as follows: Section~\ref{sec:experiments} is devoted to the description of
the experimental set-up and methods, and also to a qualitative description of the observed phase
behavior. In Section~\ref{sec:results}, the results for particle diffusion and velocity correlations
of each of the observed phases are analyzed separately and in detail. Finally, in
Section~\ref{sec:discussion} the results and final conclusions are discussed.

\section{Description of the experiments}
\label{sec:experiments}

\subsection{Setup}
\label{subsec:setup}

\begin{figure}[t!]
  \centering \includegraphics[height=0.3\textheight]{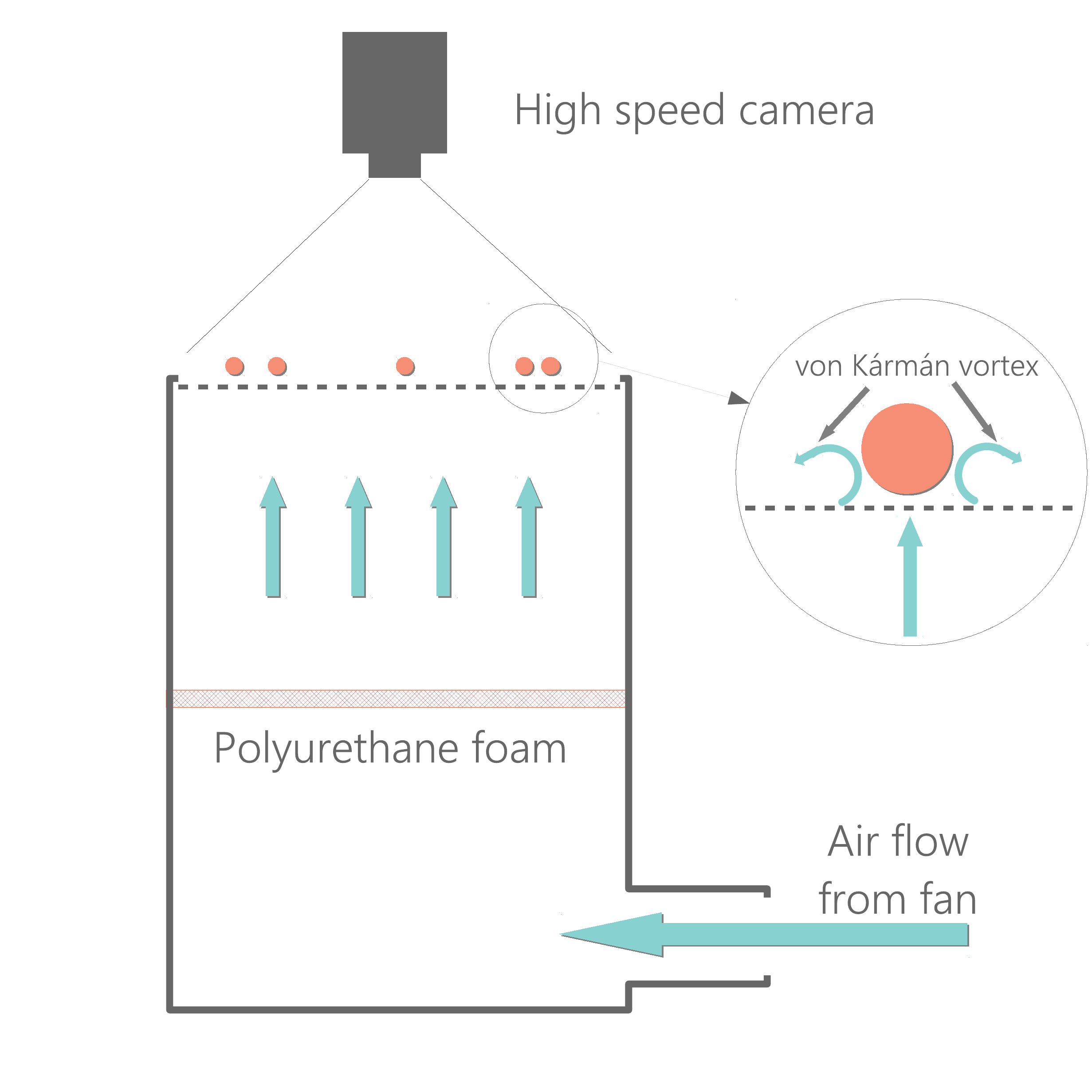}
  \caption{Sketch of the experimental set up. \label{fig:sketch} }
\end{figure}

The experimental configuration we use in this work was designed in our lab. It consists of an
air-table set-up \cite{MBF81}. In our case, it is composed by two essential parts: a) the driving
unit, that produces a stable quasi-laminar air upflow, consisting of a high power fan (SODECA
HCT-71-6T) coupled to a system of short tunnel winds; and b) the arena, which consists of a flat
metallic plate with a hexagonal lattice of perforated circular holes (of 3~mm diameter) is
surrounded by circular walls (PLA plastic) of $4.5~\mathrm{cm}$ height. The metallic plate is
carefully levelled to be horizontal (so that gravity does not enter into the dynamics if restrained
within the plate). Both parts are connected by a pair of perpendicular channels that conduct the air
released from the fan upwards to the metallic grid. See Figure~\ref{fig:sketch} for a schematic
representation of this configuration. A set of spherical particles (ping pong balls, made of ABS
plastic with mass density $0.08~\mathrm{g\,cm}^{-3}$ ) are disposed over the metallic grid. The
spherical particles are all identical, having a diameter of $\sigma=4~\mathrm{cm}$ and a mass
density $\rho=0.08~\mathrm{g\,cm^{-3}}$ (ABS plastic material). The metallic grid has a
square-shaped ($80\times80~\mathrm{cm^2}$). A (circle-shaped) plastic wall is put inside it,
centered, so that the particles are enclosed within this circular region of radius
$R = 36.25~\mathrm{cm}$.

In the between of the conducting channels there is a foam that homogenizes the upflow. This foam
helps the upflow to reach under quasi-laminar conditions, when impinges from below the set of
spherical particles. This ping-pong ball on air table configuration is inspired in a previous work
by Ohja et al., where the solution to the equation of movement of a Brownian particle (consisting on
a ping-pong ball in an air table) was found and compared to their experimental results
\cite{OLDLD04}. Fan power is carefully adjusted so that the particles never loose physical contact
with the plate and so they keep rolling over the grid, much in the same way of the aforementioned
and other previous works \cite{OLDLD04,Ojha2005,Abate2005,LGRAYV21,KMN21}.

Within the appropriate ranges of fan power, air upflow past the spheres produces turbulent vortexes
\cite{T78,vD82,OLDLD04,LGRAYV21} that yield stochastic horizontal movement to the spheres and thus
the particle dynamics (if sphere rolling is excluded) is strictly two-dimensional. For a dimensional
analysis with similar particles, please refer to the methods section in \cite{OLDLD04} and/or
supplementary material file in \cite{LGRAYV21}. As fan power is increased, the system passes through
a series of different physical configurations which are accessed through phase transitions. We have
observed phase coexistence during these transitions, for experiments in a range of values of
particle density. We characterize particle density by means of the packing fraction, that here is
defined as $\phi\equiv N\sigma^2/(4R^2)$, where $N$ is the number of particles present in the
system. For this set of experiments we used $40 \le N \le 252 $, which roughly corresponds to
packing fractions $0.12 < \phi < 0.76$. All curves presented in the manuscript result from averaging
over all of the present particles in each experiment.

We perform each experiment by setting the fan on to a constant power. Once a stationary state is
achieved, we record the particles dynamics from above by means of a high-speed camera (Phantom VEO
410L) at 250~fps. We double check afterwards that the registered interval of the experiment is in
effect under steady state conditions by plotting vs. time the relevant space-averaged magnitudes
(e.g., average kinetic energy). In any case, it is rather straightforward to ensure steady state
conditions by waiting for a time interval equivalent to several collisions per particle
\cite{MGSB99}. Since the recorded stationary section of the experiment that is recorded is always
100 s long, a large set of steady state statistical replica corresponding to recorded frames
available to process ($\approx 2\times10^4$. In this way we can achieve statistical accuracy of our
data. Data sets are obtained by processing experiments movies with a particle tracking code that we
developed specifically for this configuration. This code is composed by a series of OpenCV
\cite{opencv} and TrackPy \cite{dan_allan_2019_3492186} functions, which allow to obtain all
particle positions over the acquired images and tag each particle so that it will be tracked through
the entire movie. An exact copy of the particle tracking code (in python language) is freely
accessible \cite{ppp2_code}.

\subsection{Phase behavior}
\label{subsec:phases}

\begin{figure}[h]
  \includegraphics[height=0.33\textheight]{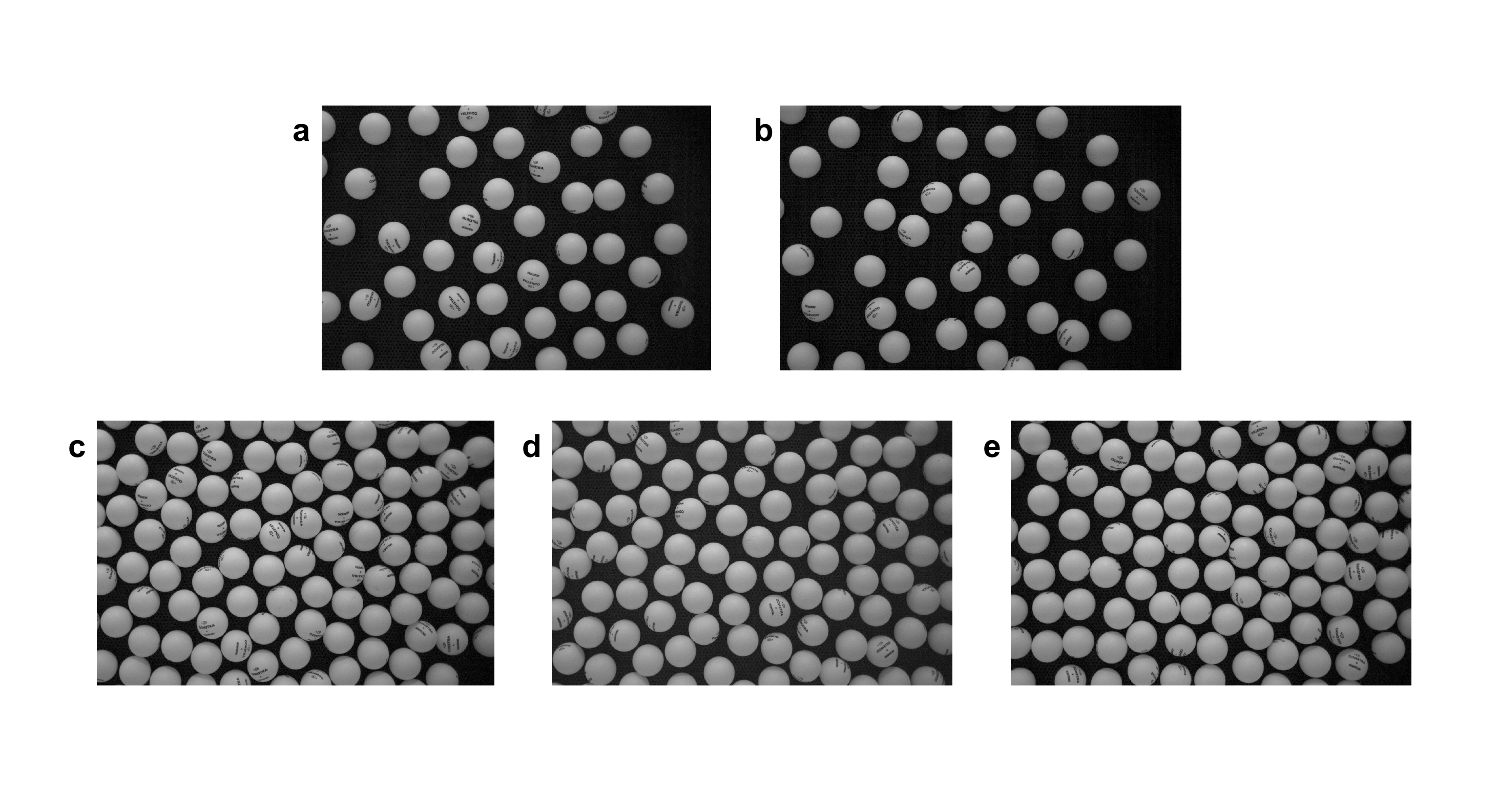}
  \caption{Snapshots of the different phase configurations observed in
    experiments. Packing fraction is $\phi=0.18$ for (a)-(b) and $\phi=0.55$ for (c)-(e). Granular
    temperatures are, in order: $T/m = [0.16,~0.74,~0.38,~0.47,~0.70]~\sigma\mathrm{^2 s^{-2}}$ for the configurations (same order): glass-arrest phase, glass-liquid, liquid, crystal, crystal-lquid.\label{fig:snapshots} }
\end{figure}

Figure~\ref{fig:snapshots} presents a series of movie snapshots displaying the different phase
states that we have detected in our experiments, for two different packing fraction values
($\phi=0.18$ for the first row, $\phi=0.55$ for the second row). For each packing fraction,
snapshots are placed in ascending order of fan power. For $\phi=0.18$ and the lowest fan power, a
subset of particles is still static, since the turbulent vortexes intensity is not strong enough so
as to overcome static friction. We denote this static phase as \textit{arrest} phase, due to its
static nature. It corresponds to the upper right corner in Figure~\ref{fig:snapshots}
(a). Interestingly, the arrest phase has been detected before in analogous configurations
\cite{OU98} and is known to develop a quasi-static ordered state that has been denoted as
\textit{collapse} phase. If current intensity is high enough, however, a subset of the spheres can
activate its thermal-like movement (that, as we discussed, is due to the turbulent vortexes
generated by the upflow past the spheres). Their movement is initially limited, so that we can
observe caging effects for these particles (bottom left section in Figure~\ref{fig:snapshots}
a). Thus, we have detected coexistence between the arrest phase and a glassy phase for the caged
moving particles (Figure~\ref{fig:snapshots} a). The arrest phase eventually disappears as particles
gradually activate, then giving rise to a pure glass phase and afterwards (at stronger upflow
intensity) to glass-liquid phase coexistence (for this coexistence, see Figure~\ref{fig:snapshots}
b, with the glass phase occupying the lower density region in the upper right corner of the
snapshot). At higher density ($\phi=0.55$, we observe consecutively: liquid phase
(Figure~\ref{fig:snapshots} c), liquid-crystal coexistence (the crystal is hexagonal, see the
developing hexagonal structure, with some defects, in bottom right corner in
Figure~\ref{fig:snapshots} d), and crystal-liquid phase (Figure~\ref{fig:snapshots} e). At this
point, if fan power is still increased, a gradual shrink of the hexagonal crystal (which
\textit{melts}). The crystal completely disappears above a threshold value of air current
intensity. At this point, only the liquid remains (again). This last stage is not represented since
they look much like the snapshots in Figure~\ref{fig:snapshots} (c),(d). In any case, grasping the
phase configuration out of these snapshots is not straightforward and for this reason we analyze in
more detail the particle trajectory structure in the next section.

\section{Results}
\label{sec:results}

\subsection{Trajectories and granular temperature field}
\label{subsec:track_T}

%%%%%%%%%%%%%%%%%%%%%%%%% TRAJECTORIES AND TEMPERATURE FIG COMMENTS %%%%%%%%%%%%%%%%%%%%%%%%%%%%%%1

In order to analyze in more detail the dynamic properties (except for the static arrest phase), we
analyze separately, for each phase, the trajectory shape, temperature field, and the properties of diffusion and
velocity autocorrelations. We define the temperature field as
$T(x,y)=(1/2)\langle v^2\rangle_{xy}$, where $\langle v^2\rangle_{xy}$ stands for the square of
particle velocity ($v$) averaged, at a given point $(x,y)$ of the system, through all measured
steady states. In this work, we use the concept of granular temperature (or simply, temperature) in
the same sense as first defined by Kanatani \cite{K79}.

Let us comment now on the phase behavior and the transitions we detected. In the first transition,
at low granular temperature, a glass is observed in coexistence with an arrest phase. By arrest
phase we refer to particles that, at very low energy input, lie still due to friction
\cite{OU98,NRTMS14}. In the second phase transitions, glass decays to a liquid-like phase,
coexisting with it as it shrinks. At even higher temperatures, a hexagonal crystallite develops , in
which we can observe in certain ranges of driving intensity a coexistence between a liquid and a
hexagonal crystal \cite{LGRAYV21}. Moreover, We have detected strong energy non-equipartition occurs
between the coexisting phases.

%%%%%%%%%%%%%%%%%%%%%%%%%%%%%%%%%%%%%%%%%%%%%%%%%%%%%%%%%%%%%%%%%%%%%%%%%%%%%%%%%%%%%%%%%%%%%%%%%%%
%%%%%%%%%%%%%%%%%%%%% TRAJECTORIES-T-g(r) FIGURE  %%%%%%%%%%%%%%%%%%%%%%%%%%%%%%%%%%%%%%%%%%%%%%%%%
%%%%%%%%%%%%%%%%%%%%%%%%%%%%%%%%%%%%%%%%%%%%%%%%%%%%%%%%%%%%%%%%%%%%%%%%%%%%%%%%%%%%%%%%%%%%%%%%%%%

\begin{figure}[H]
  \centering \includegraphics[height=0.675\textheight]{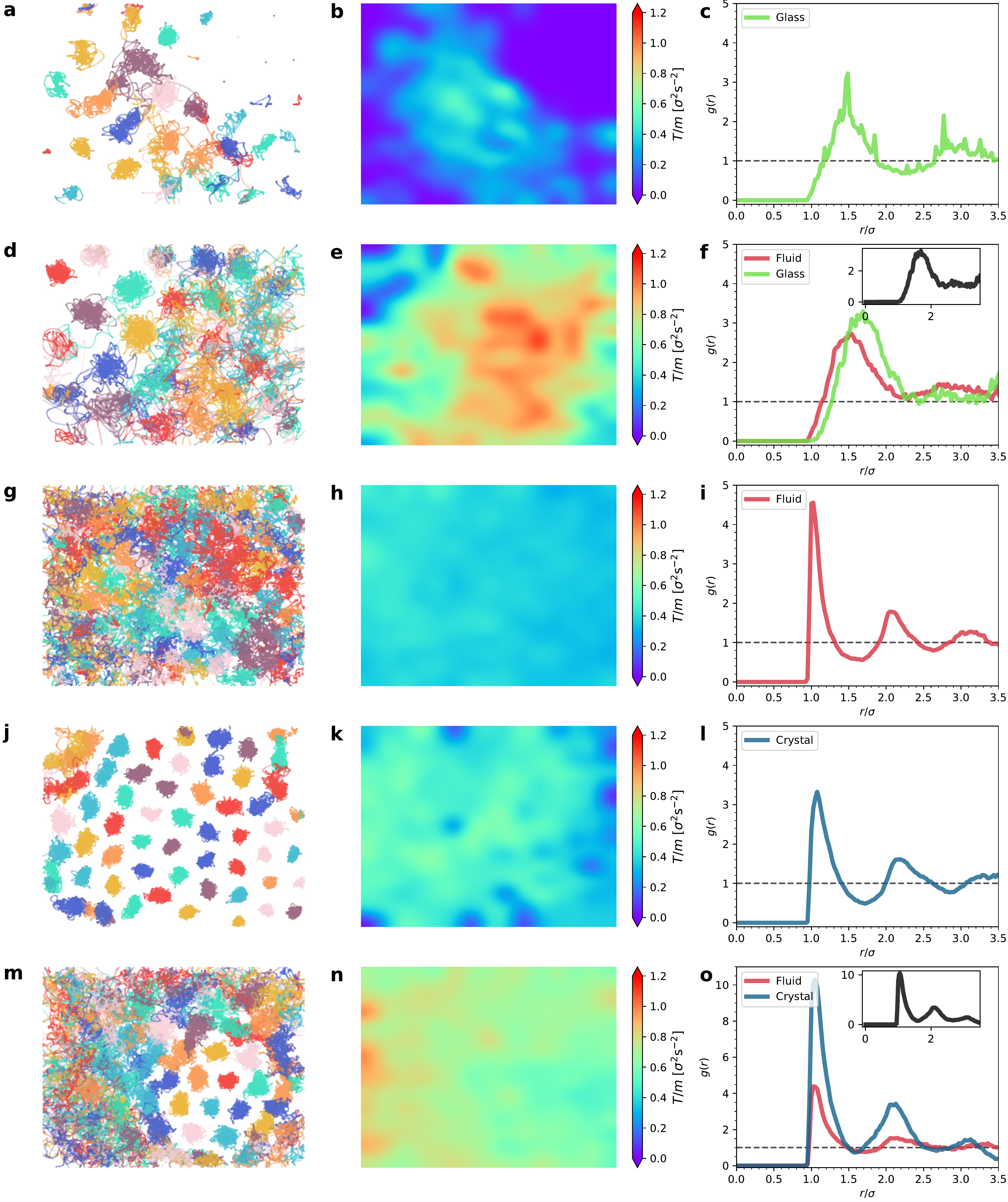}
  \caption{Phase behavior of our system, in a central region of interest. Left column represents
    particle trajectories; right column shows the corresponding granular temperature 2D fields
    ($T$). Packing fraction is $\phi=0.18$ for (a)-(f) and $\phi=0.55$ for (g)-(o). Meanwhile,
    granular temperatures for each pair of panels are, respectively:
    $T/m = [0.16,~0.74,~0.38,~0.47,~0.70]~\sigma\mathrm{^2 s^{-2}}$.
    % Granular
    % temperature is given by: $T/m$ = 0.25, 0.53 , 0.25, 0.70, 0.77 (in units of $\sigma^{2}\mathrm{s}^{-2}$) from top to bottom for both
    % columns.
    In (a)-(c) we can see phase coexistence between a glassy phase and the arrest phase at low
    density. In (d)-(f), low density but higher $T$, there is glass and liquid phase
    coexistence. (g)-(i) shows that the system is completely disordered (there is only a liquid
    phase), state that can be observed at intermediate temperatures for all densities. At higher
    densities, if the liquid is further heated (air upflow is increased), a cooler crystallite
    develops in coexistence with the liquid; the crystal grows as $T$ is increased, eventually
    occupying the entire system, as in (j)-(l). At stronger driving, the liquid tends to disappear
    and the crystal occupies the whole system, as seen in (m)-(o). \label{fig:track_T} }
\end{figure}

%%%%%%%%%%%%%%%%%%%%%%%%%%%%%%%%%%%%%%%%%%%%%%%%%%%%%%%%%%%%%%%%%%%%%%%%%%%%%%%%%%%%%%%%%%%%%%%%%%% 
%%%%%%%%%%%%%%%%%%%%%%%%%%%%%%%%%%%%%%%%%%%%%%%%%%%%%%%%%%%%%%%%%%%%%%%%%%%%%%%%%%%%%%%%%%%%%%%%%%%

These results are illustrated in Figure~\ref{fig:track_T}. This figure shows, for a representative
set of experiments, particle trajectories in the left column, 2D color maps of the granular
temperature $T(x,y)$ in the middle column and pair correlation function $g(r)$ in the right
column. The experiments shown in Figure~\ref{fig:track_T} correspond to two different densities
(low, with $\phi=0.18$ and high, with $\phi=0.55$), with each subset in ascending order of upflow
current intensity. In it, we can see the phases that consecutively appear as more energy is input
into the system. Figure~\ref{fig:track_T} (a) shows two qualitatively different types of
arrangements of particle trajectories phases: a disordered lattice of caged particle trajectories
(caged in the sense that moving particles remain close to a disordered set of fixed points), and a
disordered lattice of static particles (arrest phase). In effect, the former set of trajectories can
be identified as a glassy phase since, although particles undergo continuous stochastic movement,
caging effects are predominant \cite{Desmond2009,Rodriguez-Rivas2019} and a disordered but permanent
particle trajectory structure (lattice) can be observed.  To our knowledge, it is not very common to
find glass transitions at such low densities. With respect to the latter, it is apparent that
particles remain static during the complete 100 s experiment. From this qualitative difference
between these two phases a strong energy non-equipartition emerges. In effect, as we can see in
Figure~\ref{fig:track_T} (b), the region corresponding to the arrest phase has vanishing granular
temperature $T$ whereas for the glassy phase $T$ is clearly non-null. Note that, contrary to what
has been observed in thin layers, we have not detected a static phase that yields a hexagonally
ordered collapse phase, as in a vertically vibrated monolayer of spheres \cite{OU98}. This peculiar
arrest phase, that is present also in a vibrated granular monolayer \cite{NRTMS14}, disappears here
gradually as the upflow current is increased, to a point where we can observe two-phase coexistence
between glass-like and liquid-like phases, as in Figure~\ref{fig:track_T} (c), where the liquid
phase is observed in the region where all trajectories mix and cross each other during the
experiment, in contrast with the disordered pattern of localized trajectories that is visible in the
upper left corner. As usual \cite{Desmond2009}, the corresponding $g(r)$ curves for the glass phases
(see green curves in Figure~\ref{fig:track_T} c,f) do not help to distinguish the glass phase from
the liquid phase. We will have to resort to the peculiar dynamic properties of the glass for that
matter (mean squared displacement and velocity autocorrelations). As we can see in
Figure~\ref{fig:track_T} (e), energy non-equipartition is strong here again, with the glass phase
being noticeably cooler. At higher density (packing fraction $\phi=0.55$), we observe,
consecutively, a monophase liquid-like system (Figure~\ref{fig:track_T}~f-h); a hexagonal crystal
phase (Figure~\ref{fig:track_T}~i-k); and a two-phase system, with a liquid coexisting with a
hexagonal lattice (Figure~\ref{fig:track_T}~i-j). With respect to the corresponding behavior of
their pair correlation function, $g(r)$, the curves for the liquid and the crystal are noticeably
different, with the curves for the crystal having stronger secondary peaks, as usual
\cite{OU05}. With respect to $T(x,y)$, notice also that non-equipartition is also present in the
case of the liquid-crystal two phase system, with the crystal colder than the liquid
(Figure~\ref{fig:track_T} h).

%%%%%%%%%%%%%%%%%%%%%%%%%%%%%%%%%%%%%%%%%%%%%%%%%%%%%%%%%%%%%%%%%%%%%%%%%%%%%%%%%%%%%%%%%%%%%%%%%%%

%%%%%%%%%%%%%%%%%%%%%%%%%%%%%%%%%%%%%%%%%%%%%%%%%%%%%%%%%%%%%%%%%%%%%%%%%%%%%%%%%%%%%%%%%%%%%%%%%%%
%%%%%%%%%%%%%%%%%%%%% MSD FIGURE, low density     %%%%%%%%%%%%%%%%%%%%%%%%%%%%%%%%%%%%%%%%%%%%%%%%%
%%%%%%%%%%%%%%%%%%%%%%%%%%%%%%%%%%%%%%%%%%%%%%%%%%%%%%%%%%%%%%%%%%%%%%%%%%%%%%%%%%%%%%%%%%%%%%%%%%%

\begin{figure}[t!]
  \includegraphics[width=0.95\textwidth]{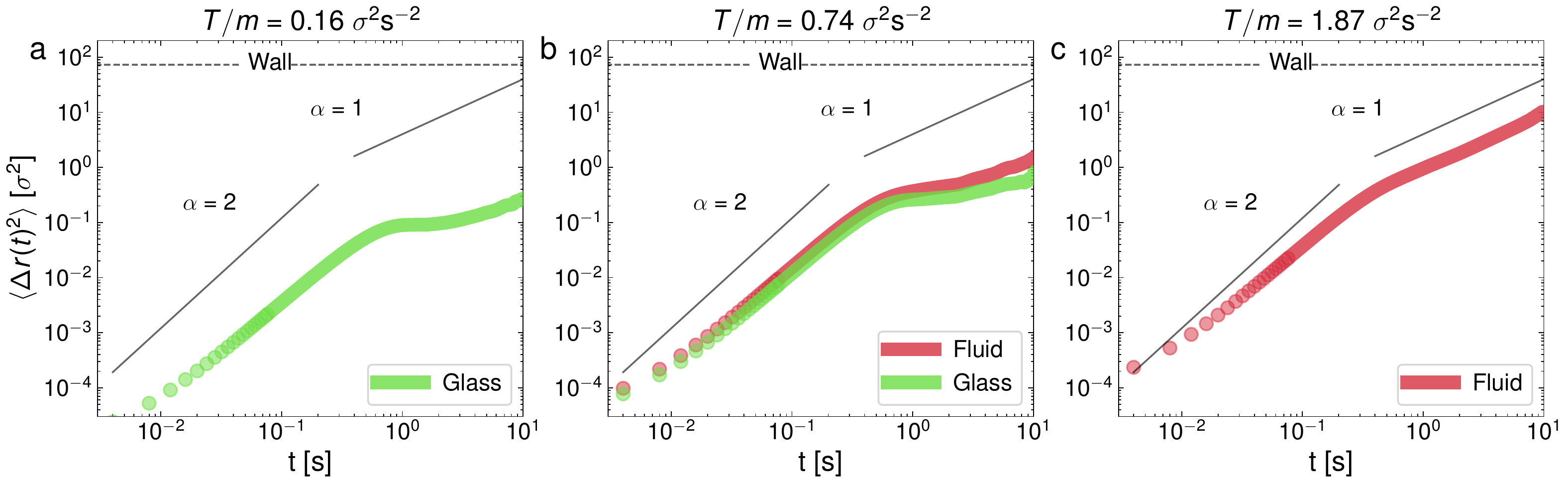}
  \caption{Log-Log representation of the mean squared displacement for three representative
    experiments of $\phi = 0.18$, the central panel corresponds to the case presented in
    Figure~\ref{fig:track_T}(c)-(d).}
\label{fig:msd}
\end{figure} 

%%%%%%%%%%%%%%%%%%%%%%%%%%%%%%%%%%%%%%%%%%%%%%%%%%%%%%%%%%%%%%%%%%%%%%%%%%%%%%%%%%%%%%%%%%%%%%%%%%% 
%%%%%%%%%%%%%%%%%%%%%%%%%%%%%%%%%%%%%%%%%%%%%%%%%%%%%%%%%%%%%%%%%%%%%%%%%%%%%%%%%%%%%%%%%%%%%%%%%%%

%%%%%%%%%%%%%%%%%%%%%%%%%%%%%%%%%%%%%%%%%%%%%%%%%%%%%%%%%%%%%%%%%%%%%%%%%%%%%%%%%%%%%%%%%%%%%%%%%%%
%%%%%%%%%%%%%%%%%%%%% MSD FIGURE, high density     %%%%%%%%%%%%%%%%%%%%%%%%%%%%%%%%%%%%%%%%%%%%%%%%%
%%%%%%%%%%%%%%%%%%%%%%%%%%%%%%%%%%%%%%%%%%%%%%%%%%%%%%%%%%%%%%%%%%%%%%%%%%%%%%%%%%%%%%%%%%%%%%%%%%%

\begin{figure}[t!]
  \includegraphics[width=0.95\textwidth]{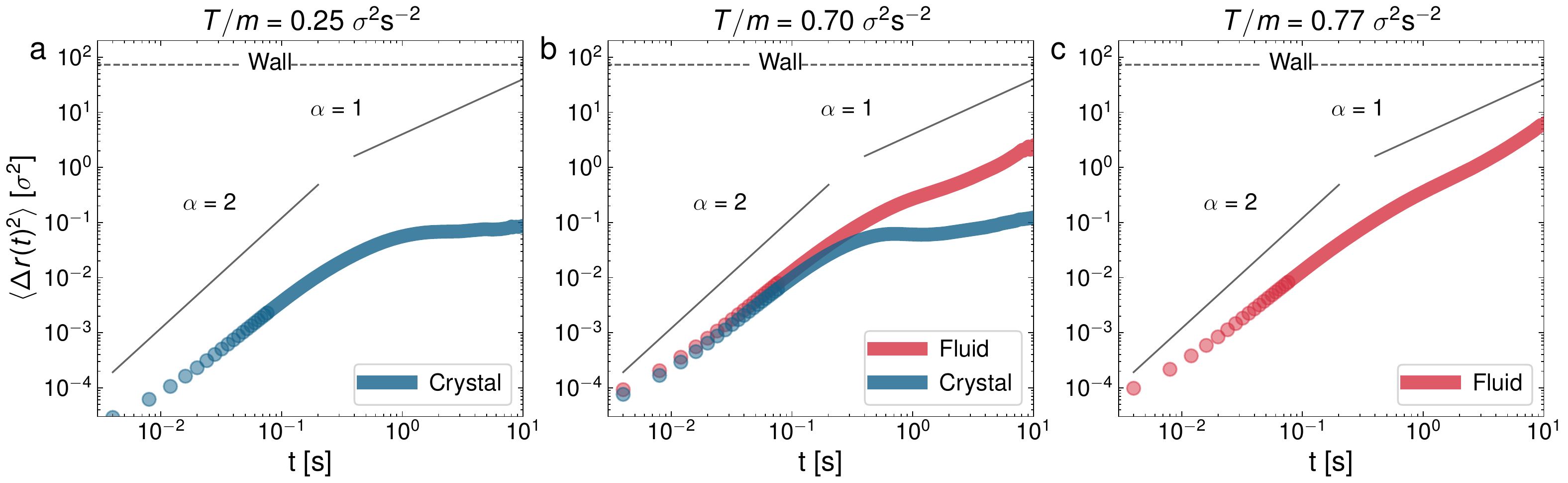}
  \caption{Log-Log representation of the mean squared displacement for three representative
    experiments of $\phi = 0.55$. The central panel corresponds to the case presented in
    Figure~\ref{fig:track_T}(i)-(j).} \label{fig:msd0}
\end{figure}

%%%%%%%%%%%%%%%%%%%%%%%%%%%%%%%%%%%%%%%%%%%%%%%%%%%%%%%%%%%%%%%%%%%%%%%%%%%%%%%%%%%%%%%%%%%%%%%%%%% 
%%%%%%%%%%%%%%%%%%%%%%%%%%%%%%%%%%%%%%%%%%%%%%%%%%%%%%%%%%%%%%%%%%%%%%%%%%%%%%%%%%%%%%%%%%%%%%%%%%%

%%%%%%%%%%%%%%%%%%%%%%%%%%%%%%%%%%%%%% MSD FIGURE COMMENTS %%%%%%%%%%%%%%%%%%%

Overall, the fact that non-equipartition is noticeably present in all two-phase system configuration
denotes that each phase has its own peculiar dynamics. In particular, this can be an indication that
the diffusion process in each phase might have different scales and behavior \cite{MJCB14}. In this
sense, note the structure of Brownian trajectories in each phase is very different, as trajectories
in the left column of Figure~\ref{fig:track_T} show. For this reason, by identifying first which
trajectories belong to each phase in all experiments (with the aid of the average coordination
number for each particle, as extracted from the corresponding Voronoi tesselations \cite{LGRAYV21}),
we have computed the diffusion coefficient for each phase. This allows us for tracking the mean
squared displacement (MSD) for each phase independently. Figures~\ref{fig:msd},~\ref{fig:msd0} show
the evolution of the ensemble mean squared displacements, that in 2D can be defined as

\begin{equation}
  \label{eq:msd}
\langle\Delta r(t)^2\rangle\equiv\langle\Delta x(t)^2 + \Delta y(t)^2\rangle,
\end{equation}
where $\Delta x(t)^2\equiv (1/\mathcal{N}(t))\sum_{\{t_0\}}[x(t+t_0)-x(t_0)]^2$ (and analogously for
$\Delta y(t)^2$). For each lag time, and under steady state conditions, the squared displacements
$\Delta x(t)^2, \Delta y(t)^2$ can be obtained from averages over the $\mathcal{N}(t)$ available
initial times $t_0$ (basically, $\mathcal{N}=N\times N_\mathrm{st,frames}$ where
$N_\mathrm{st,frames}$ is the number of images taken under steady state conditions, in our case,
$N_\mathrm{st,frames}\approxeq 25000$). The diffusion coefficient can be obtained from the MSD,
since most commonly the following relation is fulfilled \cite{MJCB14,LGRAYV21}

\begin{equation}
  \label{eq:diff}
  \langle\Delta r(t)^2\rangle=(4D)t^\alpha,
\end{equation}
where $\alpha$ is a constant usually called diffusive exponent. (When Equation~\ref{eq:diff} is not
fulfilled, diffusion is said to be anomalous \cite{MJCB14}). Trivially from \eqref{eq:diff}, the
diffusive exponent $\alpha$ corresponds to the slope of the diffusive part, in a Log-Log
representation, of the MSD vs. lag time, as in the curves displayed in
Figures~\ref{fig:msd},~\ref{fig:msd0}. By diffusive part we mean, as usual, the part of the MSD
vs. time curve that is after the ballistic regime, which should always have $\alpha=2$
\cite{MJCB14}. As a guide to the eye, the ballistic ($\alpha=2$) and normal diffusion ($\alpha=1$)
diffusion values were indicated inside each panel in Figures~\ref{fig:msd},~\ref{fig:msd0}. We have
found however that the slope in the diffusive part of the MSD curves is not constant in some cases,
and it may be said that diffusion is in these cases anomalous, which is in agreement with the
experimental observations a previous analysis \cite{KMN21}. Thus, the computed diffusive exponents
$\alpha$ for these cases, and the corresponding slope (and thus, diffusion coefficient) from
Equation~\ref{eq:diff} represent an averaged value.

In Figure~\ref{fig:msd}, at low packing fraction ($\phi=0.18$) we can see the MSD for the following
cases: observed; glass (a), glass-liquid (b), liquid (c); whereas in Figure~\ref{fig:msd0} we can
see the cases: only crystal (a), crystal-liquid (b) and only liquid(c). It is very apparent that the
behavior of the MSD for each phase is very different. In particular, the monophase glassy
configuration (Figure~\ref{fig:msd} a) presents a MSD with a local maximum at the end of the
ballistic regime, after which it presents a characteristic curvature in the diffusive part of the
curve, which is besides strongly subdiffusive. The MSD behaviour of the glass-like phase is thus
characterized by a short plateau in the MSD followed by an increase (when particles escape the
current "caging" area and move to a new location. At higher $T$, in Figure~\ref{fig:msd} (b), we can
see the glass-liquid coexistence. In this case, the emerging liquid phase is still weakly
subdiffusive (although with a clearly faster MSD, if compared to the companion glass). This can be
attributed to a liquid structure that is still in development as the glass shrinks in size. In
Figure~\ref{fig:msd} (c), the system with only liquid phase shows already a normal diffusion
scenario. By contrast, Figure~\ref{fig:msd0} (a), at higher packing fraction ($\phi=0.55$), shows a
single crystal configuration, with the diffusive part of the MSD close to stagnation (zero time
growth of the MSD); i.e., the dynamics is very strongly subdiffusive, as an evidence of crystalline
diffusion \cite{RIS06}. In the case of crystal-liquid coexistence, Figure~\ref{fig:msd0} (b), the
less disordered phase (glass) clearly undergoes subdiffusion, whereas the liquid has normal
diffusion. Normal diffusion can also be seen in Figure~\ref{fig:msd0} (c), where the single liquid
phase is recovered. As we said before, and is still worth to remark here again, an important and
surprising result is the confirmation, in view of the MSD behavior illustrated in
Figure~\ref{fig:msd}, of a glassy transitions with clear caging processes at low densities, when in
general these processes are observed (to the best of our knowledge) in dense granular fluids
\cite{KSZ10}. This result may be the outcome of an effective potential developed by the interaction
between the spherical balls through the intermediate air flow. In the same way, the crystal appears
at unusually low densities, in comparison with the case of hard particles \cite{OU05}.

%%%%%%%%%%%%%%%%%%%%%%%%%%%%%%%%%%%%%%%%%%%%%%%%%%%%%%%%%%%%%%%%%%%%%%%%%%%%%%%%%%%%%%%%%%%%%%%%%%%

\subsection{Diffusion coefficient}
\label{subsec:diff}

\begin{figure}[t!]
  \includegraphics[width=0.471\textwidth]{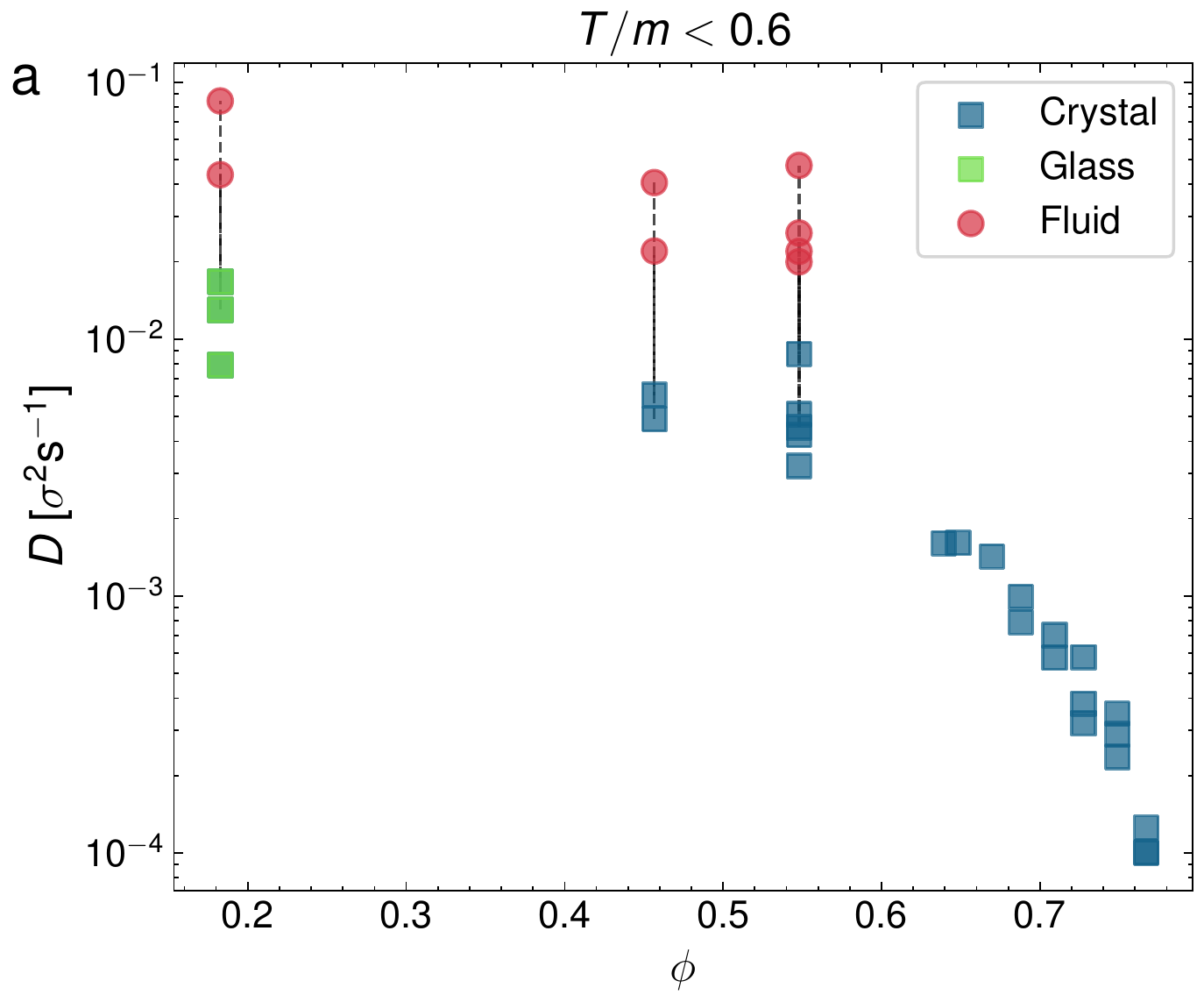}
  \includegraphics[width=0.471\textwidth]{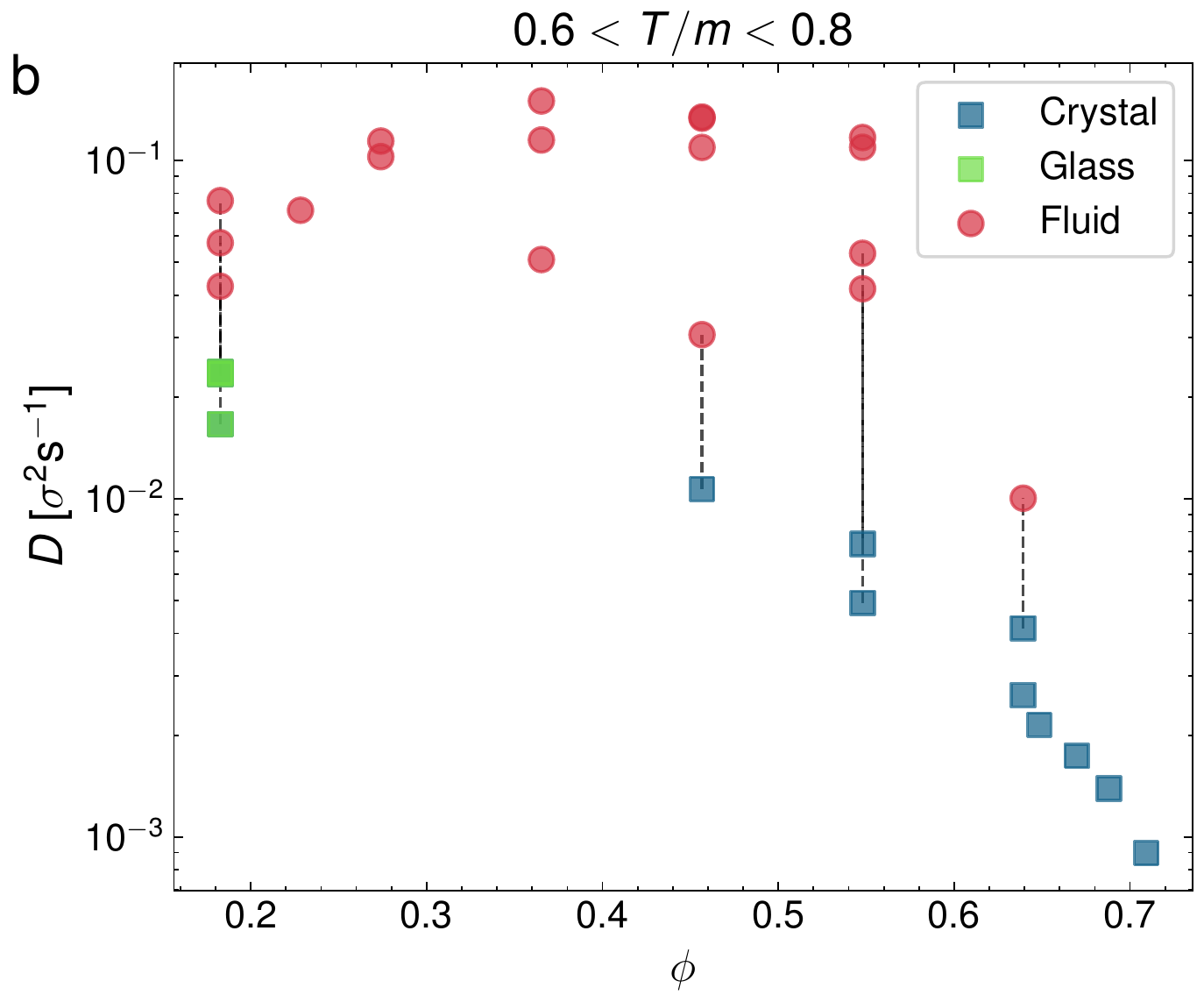}
  \includegraphics[width=0.471\textwidth]{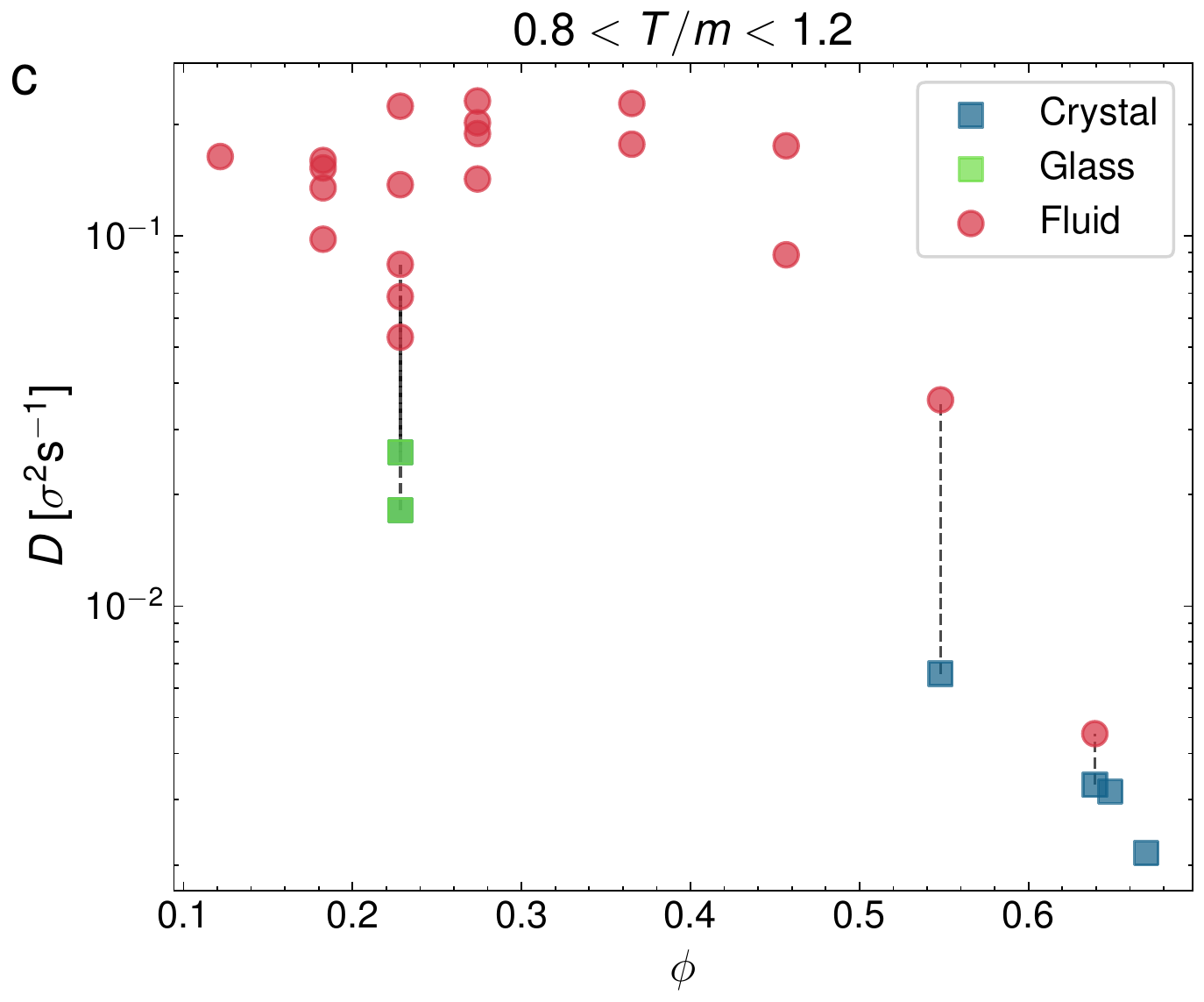}
  \includegraphics[width=0.471\textwidth]{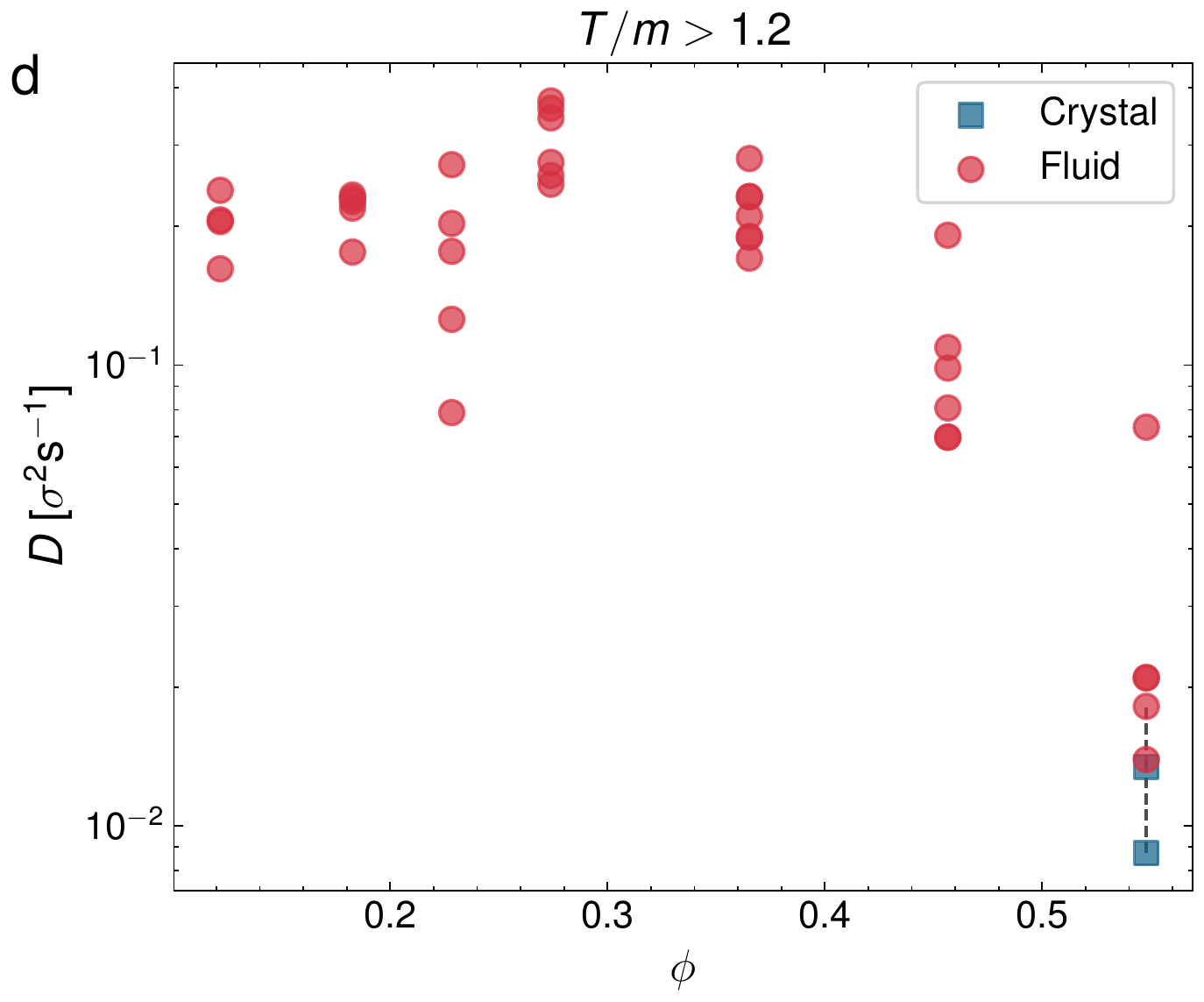}
\caption{Diffusion coefficient $D$ vs. packing fraction $\phi$ divided in four panels by the overall granular temperature
of each experiment.\label{fig:D_vs_phi}}
\end{figure}

\begin{figure}[t]
  \includegraphics[width=0.33\textwidth]{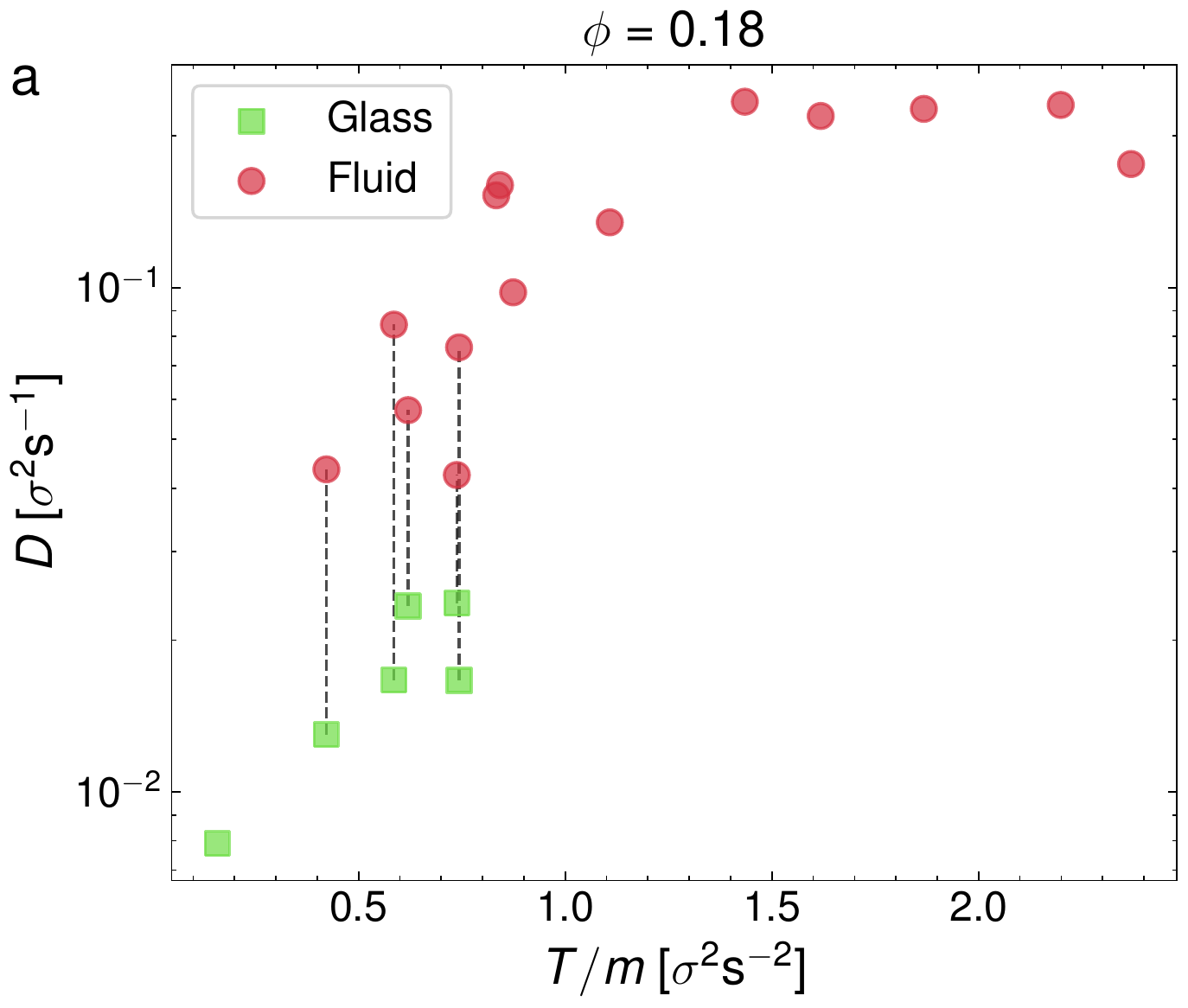}
  \includegraphics[width=0.33\textwidth]{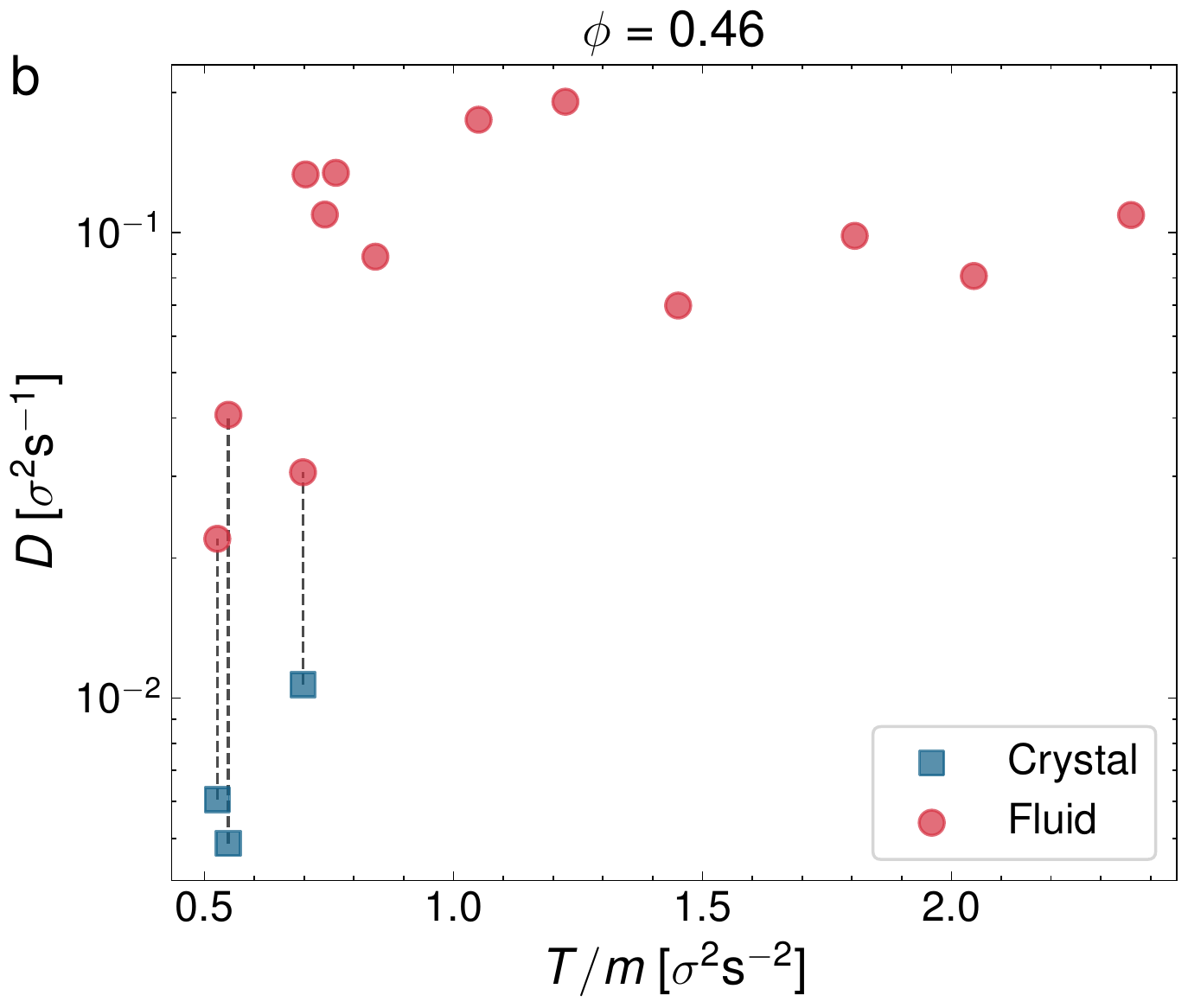}
  \includegraphics[width=0.33\textwidth]{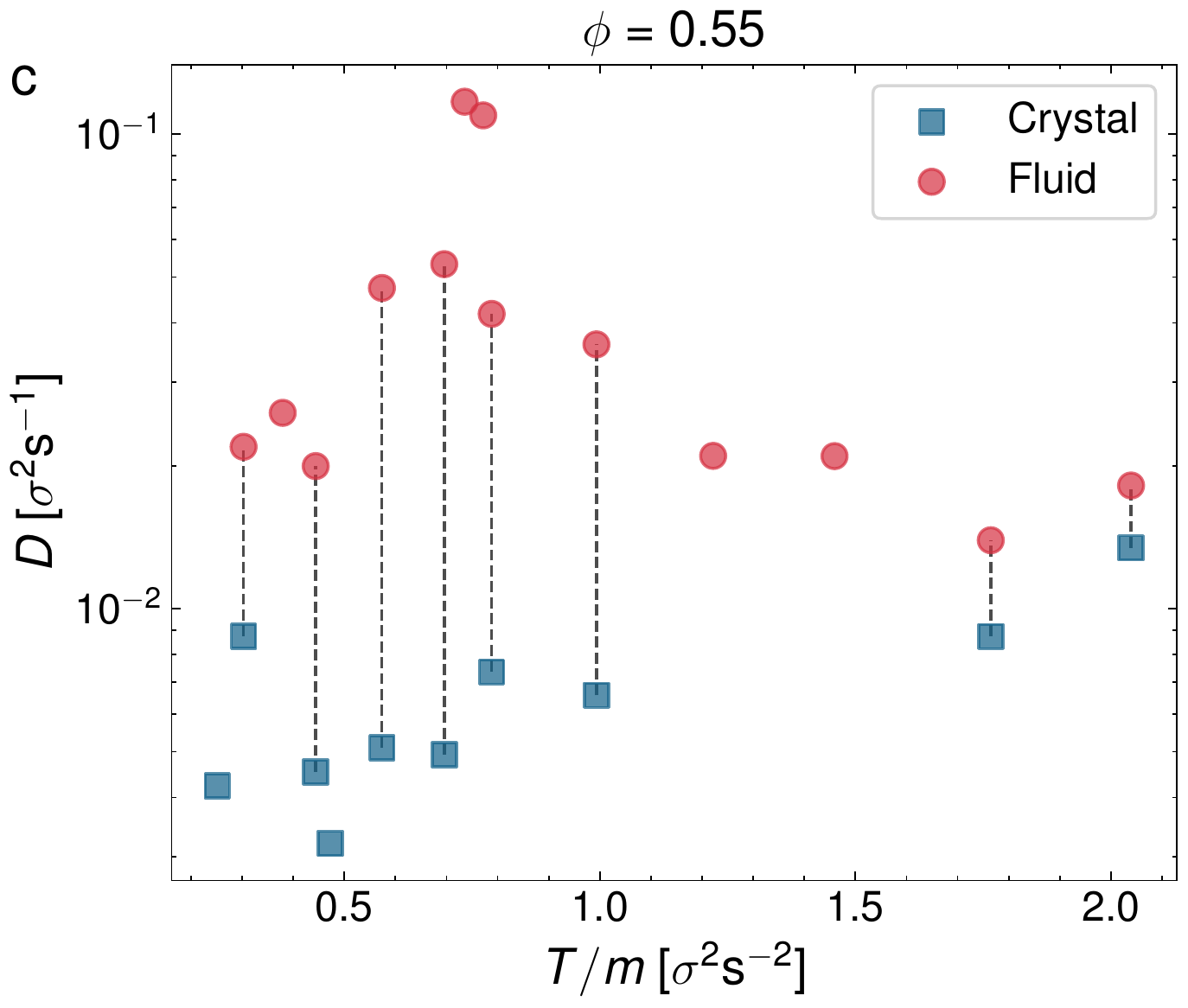}
  \caption{Average diffusion coefficients represented against granular temperature for three
    different packing fractions. Each point corresponds to an experiment; where coexistence is
    visible, we have split $D$ into two different points, for the fluid (red) and crystal/glass
    phase (blue/green). \label{fig:D_vs_T}}
\end{figure}

%%%%%%%%%%%%%%%%%%%%%%%%%%%%%%%%%% DIFFUSION COEFFS COMMENTS %%%%%%%%%%%%%%%%%%%%%%%%%%%%%%%%%%%%%
From the results in Figures~\ref{fig:msd},~\ref{fig:msd0}, we may conclude that the evolution of the
MSD for each phase is qualitatively very different, which confirms our identification of the
different observed phases, as previously discussed. Next, we compute the diffusion coefficient
separately for each phase in this section. We represent in two figures our measurements of the
diffusion coefficient. In Figure~\ref{fig:D_vs_phi}, $D$ is represented vs. packing fraction, for a
series of experiments in different ranges of $T$:
$T/m<0.6\; \sigma^2/\mathrm{s}^2;\; 0.6~\sigma^2/\mathrm{s}^2< T/m < 0.8~\sigma^2/\mathrm{s}^2;\;
0.8~\sigma^2/\mathrm{s}^2< T/m <1.2~\sigma^2/\mathrm{s}^2$ and $T/m> 1.2~\sigma^2/\mathrm{s}^2$,
whereas in Figure~\ref{fig:D_vs_T}, we plot $D$ vs. $T$ for three representative packing fraction
values ($\phi=0.18; 0.46; 0.55$).

Figure~\ref{fig:D_vs_phi} highlights the diffusive stages of the different phase configurations,
including those with phase coexistence (the coexisting phases are here joined with dashed vertical
lines). As we can see, at $T/m<0.6~\sigma^2/\mathrm{s}^2$ (top left panel), the diffusion
coefficient tends in general to decrease for increasing $\phi$. Moreover, only glass or liquid
phases are visible at very low $T$, with the liquid coexisting with the glass at low packing
fractions whereas at intermediate packing fractions we find crystal-liquid coexistence and at larger
$\phi$ only the crystal is detected, in this case with the lowest $D$ values. At higher intermediate
temperatures (at $0.6~\sigma^2/\mathrm{s}^2< T/m < 0.8~\sigma^2/\mathrm{s}^2$, in top right panel;
and at $0.8~\sigma^2/\mathrm{s}^2< T/m <1.2~\sigma^2/\mathrm{s}^2$, bottom left) we can see the
glass-liquid at low density again the effects of larger $T$ cause the withdrawal of the
crystal-liquid coexistence at intermediate $\phi$, leaving the liquid (red symbols) alone. Again, at
higher $\phi$, crystal-liquid and crystal are detected.  Finally, in the largest range of values of
$T$, it is apparent that only the liquid is observed (except for a configuration with the densest
system we used) and that in this regime the diffusion coefficient is nearly constant with respect to
packing fraction, except for a steep decay at large $\phi$ (where the only two cases of coexistence
with a crystal are here observed). It is also interesting to note that an extrapolation of the curve
averaged by the crystalline states extends to the low-density glass transition zones.  In summary,
$D$ tends to decrease for denser systems, except at very high $T$, where it tends to keep
approximately constant. 

Now, in Figure~\ref{fig:D_vs_T}, which represents $D$ vs. $T$, summarizes well the quantitative
differences in the diffusion coefficient for the three phases (glass, liquid, crystal), together
with the ranges of coexistence of glass and crystal with the liquid phase. Overall, liquid
predominates at low and moderate density (left and center panels), whereas glass and crystal
predominate at very low and high density respectively. It can also be observed that both glass and
crystal are less diffusive than the liquid, as it was to be expected, with the crystal having the
lowest values, systematically, of the diffusion coefficient.

It is important to remark here that Figures~\ref{fig:D_vs_phi},~\ref{fig:D_vs_T} are plenty with
evidences of finite differences of the diffusion coefficient in the same system state, as a strong
experimental proof of discontinuous phase transition and phase coexistence. Moreover, we have not
observed in any case a continuously changing diffusion coefficient between the different phases.

%%%%%%%%%%%%%%%%%%%%%%%%%%%%%%%%%%%%%%%%%%%%%%%%%%%%%%%%%%%%%%%%%%%%%%%%%%%%%%%%%%%%%%%%%%%%%%%%%%%
%%%%%%%%%%%%%%%%%%%%% DIFFUSION COEFFICIENT FIGURE, vs density     %%%%%%%%%%%%%%%%%%%%%%%%%%%%%%%%
%%%%%%%%%%%%%%%%%%%%%%%%%%%%%%%%%%%%%%%%%%%%%%%%%%%%%%%%%%%%%%%%%%%%%%%%%%%%%%%%%%%%%%%%%%%%%%%%%%%

%%%%%%%%%%%%%%%%%%%%%%%%%%%%%%%%%%%%%%%%%%%%%%%%%%%%%%%%%%%%%%%%%%%%%%%%%%%%%%%%%%%%%%%%%%%%%%%%%%%
%%%%%%%%%%%%%%%%%%%%%%%%%%%%%%%%%%%%%%%%%%%%%%%%%%%%%%%%%%%%%%%%%%%%%%%%%%%%%%%%%%%%%%%%%%%%%%%%%%%

%%%%%%%%%%%%%%%%%%%%%%%%%%%%%%%%%%%%%%%%%%%%%%%%%%%%%%%%%%%%%%%%%%%%%%%%%%%%%%%%%%%%%%%%%%%%%%%%%%%
%%%%%%%%%%%%%%%%%%%%% DIFFUSION COEFFICIENT FIGURE, vs temperature    %%%%%%%%%%%%%%%%%%%%%%%%%%%%%
%%%%%%%%%%%%%%%%%%%%%%%%%%%%%%%%%%%%%%%%%%%%%%%%%%%%%%%%%%%%%%%%%%%%%%%%%%%%%%%%%%%%%%%%%%%%%%%%%%% 

%%%%%%%%%%%%%%%%%%%%%%%%%%%%%%%%%%%%%%%%%%%%%%%%%%%%%%%%%%%%%%%%%%%%%%%%%%%%%%%%%%%%%%%%%%%%%%%%%%%
%%%%%%%%%%%%%%%%%%%%%%%%%%%%%%%%%%%%%%%%%%%%%%%%%%%%%%%%%%%%%%%%%%%%%%%%%%%%%%%%%%%%%%%%%%%%%%%%%%%

\begin{figure}[ht!]
  \centering\includegraphics[width=0.45\textwidth]{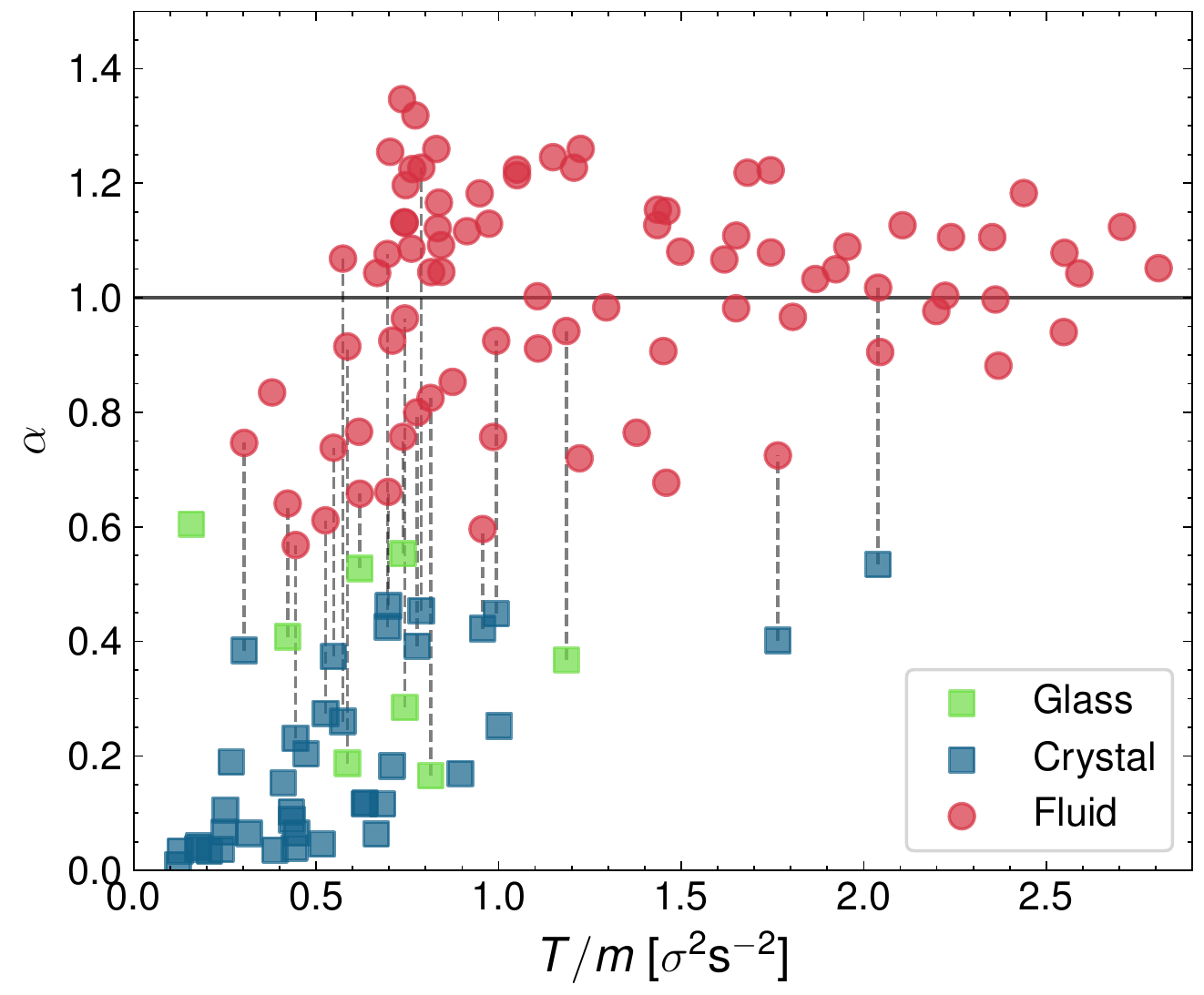}
\caption{Diffusive exponent represented against granular temperature for all experiments. It has
  been calculated by averaging the logarithmic slope of the MSD in the [3-6] s range. Each point
  corresponds to an experiment; where coexistence is visible, we have split $D$ into two different
  points, for the fluid (red) and crystal/glass phase (blue/green). \label{fig:alpha}}
\end{figure}

%%%%%%%%%%%%%%%%%%%%%%%%%%% ALPHA FIGURE  COMMENTS %%%%%%%%%%%%%%%%%%%%%%%%%%%%%%%%%%%%%%%%%%%%%%%%%
As we mentioned before, previously to computing the diffusion coefficient we determine the diffusive
exponent $\alpha$ as defined by eq. (\ref{eq:diff}), and whose value defines if the system is under
super-diffusion ($\alpha>1$), sub-diffusion ($\alpha<1$) or normal diffusion ($\alpha=1$)
\cite{LVR22,MJCB14}. So, we plot in Figure~\ref{fig:alpha} the measurement of $\alpha$ for all the
performed experiments altogether. They are represented as a function system granular temperature $T$
for all the particle densities combined (here, represented in the form of packing fraction
$\phi$). Red points signal the liquid phase diffusive exponents, green stands for the glass phase
and blue for the crystal. Note the logical order of $\alpha$ by phases, with the crystal having the
lowest values, the glass in the intermediate region and the liquid having the largest
$\alpha$. Also, as we can see, the crystal and also glass phases are very subdiffusive. The liquid
however can be either weakly sub-diffusive or weakly superdiffusive. Superdiffusive values
($\alpha>1$) are reached after phase coexistence has vanished; i.e., the pure liquid phase tends to
be superdiffusive, which we think is an indication again of repulsive forces between the
particles. Specially at not large densities, repulsion between particles may aide the particles
diffuse in between neighboring particles, thus enhancing diffusion.

%%%%%%%%%%%%%%%%%%%%%%%%%%%%%%%%%%%%%%%%%%%%%%%%%%%%%%%%%%%%%%%%%%%%%%%%%%%%%%%%%%%%%%%%%%%%%%%%%%%
%%%%%%%%%%%%%%%%%%%%% DIFFUSIVE EXPONENT FIGURE   %%%%%%%%%%%%%%%%%%%%%%%%%%%%%%%%%%%%%%%%%%%%%%%%%
%%%%%%%%%%%%%%%%%%%%%%%%%%%%%%%%%%%%%%%%%%%%%%%%%%%%%%%%%%%%%%%%%%%%%%%%%%%%%%%%%%%%%%%%%%%%%%%%%%%

%%%%%%%%%%%%%%%%%%%%%%%%%%%%%%%%%%%%%%%%%%%%%%%%%%%%%%%%%%%%%%%%%%%%%%%%%%%%%%%%%%%%%%%%%%%%%%%%%%% 
%%%%%%%%%%%%%%%%%%%%%%%%%%%%%%%%%%%%%%%%%%%%%%%%%%%%%%%%%%%%%%%%%%%%%%%%%%%%%%%%%%%%%%%%%%%%%%%%%%%

\subsection{Velocity autocorrelations}
\label{subsec:autocorrelations}

%%%%%%%%%%%%%%%%%%%%%%% discussion on  Velocity autocorrelations  %%%%%%%%%%%%%%%%%%%%%%%%%%%%%%%%%

\begin{figure}[ht]
  \includegraphics[width=0.99\textwidth]{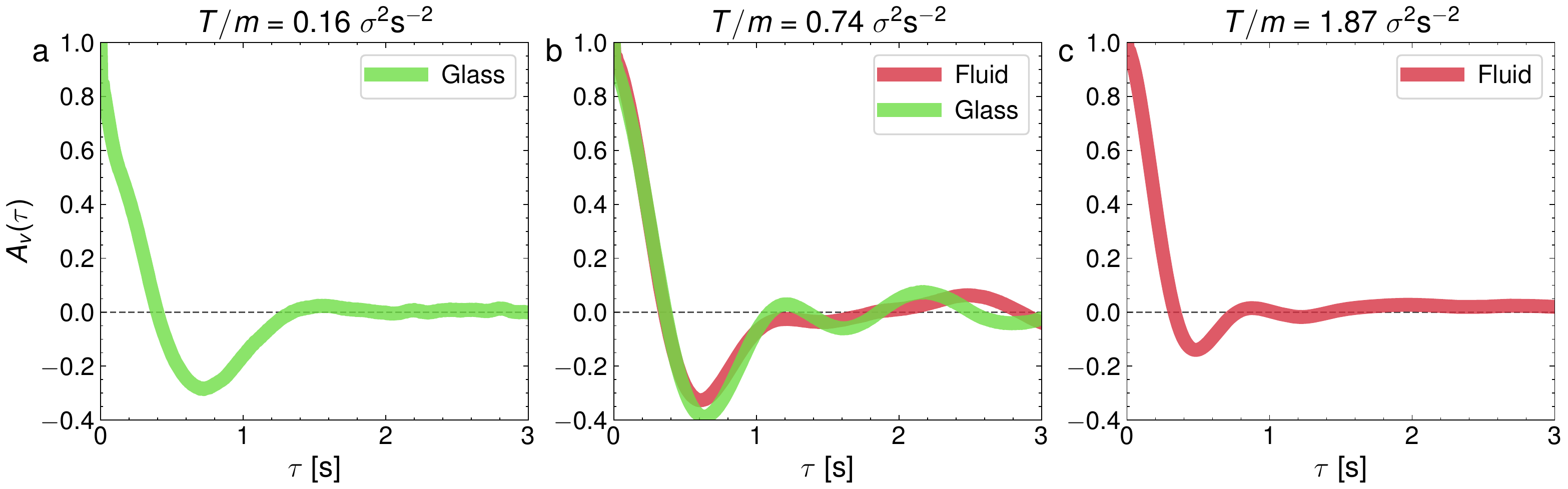}
\caption{Normalized velocity autocorrelation for three different temperatures at $\phi = 0.18$. They correspond to the cases presented for the MSD in Figure~\ref{fig:msd}. \label{fig:A_dilute}}
\end{figure}

\begin{figure}[ht]
  \includegraphics[width=0.99\textwidth]{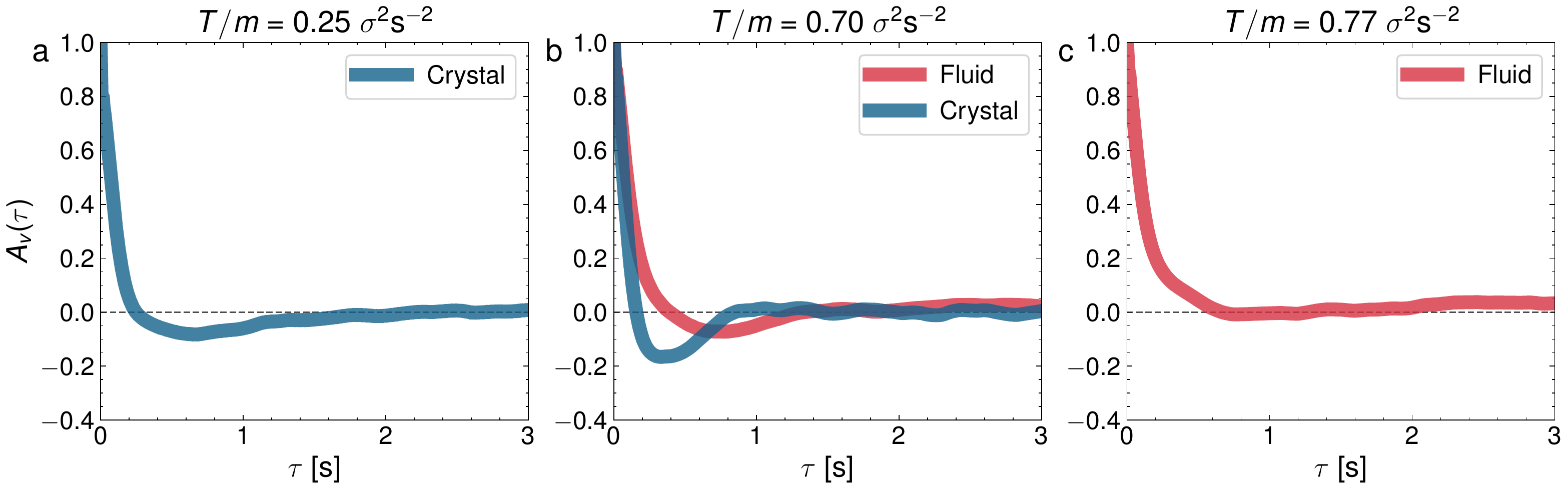}
\caption{Normalized velocity autocorrelation for three representative experiments of $\phi = 0.55$. They correspond to the cases presented for the MSD in Figure~\ref{fig:msd0}. \label{fig:A_dense}}
\end{figure}

We also represented the velocity autocorrelation function, computing its trend for glass, liquid and
crystal. Velocity autocorrelations provide information on the dynamics of particle collision, in
particular on the statistical relation between pre-collisional and post-collisional velocities. We
define the velocity autocorrelation function at lag time $\tau$ as usual \cite{Melby2005}

\begin{equation}
    A_{\mathbf{v}}(\tau) = \frac{\langle\mathbf{v}(t) \cdot\mathbf{v}(t+\tau)\rangle}{\langle\mathbf{v}(t)\cdot\mathbf{v}(t)\rangle},
\label{eq:autocorrelation}
\end{equation} where here $\langle\dots\rangle$ stands for ensemble averaging over all
steady states at initial times $t_0$. Figure~\ref{fig:A_dilute} represents velocity correlations $A_{\mathbf{v}}(\tau)$ in
the glass-liquid transition and Figure~\ref{fig:A_dense} represents $A_{\mathbf{v}}(\tau)$ for the
crystal-liquid transition. It is to be noted that particles in the glass phase (left panel in
Figure~\ref{fig:A_dilute}) show strong velocity anticorrelations at early times
($A_{\mathbf{v}}(\tau<1)<0$) and that these anticorrelations are 
transmitted to the coexisting liquid (center panel of Figure~\ref{fig:A_dilute}). Surprisingly as
well, the depth of the anticorrelation well is increased in the glass-liquid two-phase system, with
respect to the pure glass (left panel). Furthermore, the liquid remains anticorrelated at $\tau<1$
even when the glass has disappeared at high $T$. By contrast, the pure liquid phase does not
display autocorrelations in the crystal-liquid transition (right panel in
Figure~\ref{fig:A_dense}), as well as an increase in the time required for the autocorrelation 
function to cancel for the first time. However, there are weaker anticorrelations in the pure crystal (left
panel in Figure~\ref{fig:A_dense}) and crystal-liquid two-phase state, and a very short time for 
the first cancellation of the autocorrelation function, typical of the crystal phase. Let us remark here that
the right panels in
Figures~\ref{fig:A_dilute},~\ref{fig:A_dense} combined reveal that the liquid phase has a variety of
internal behaviors. This variety of behavior is closely related to the occurrence of the glass 
transition at low densities, because as mentioned before, at low densities there is a repulsive 
interaction between the particles mediated by the upward air flow (as if they had a soft core 
with a diameter greater than that of the balls), and when the density is increased, this effective 
potential does not prevent the direct collision between the spherical balls.

%%%%%%%%%%%%%%%%%%%%%%%%%%%%%%%%%%%%%%%%%%%%%%%%%%%%%%%%%%%%%%%%%%%%%%%%%%%%%%%%%%%%%%%%%%%%%%%%%%%

%%%%%%%%%%%%%%%%%%%%%%%%%%%%%%%%%%%%%%%%%%%%%%%%%%%%%%%%%%%%%%%%%%%%%%%%%%%%%%%%%%%%%%%%%%%%%%%%%%%
%%%%%%%%%%%%%%%%%%%%% VELOCITY AUTOCORRELATIONS FIGURE, low density   %%%%%%%%%%%%%%%%%%%%%%%%%%%%%
%%%%%%%%%%%%%%%%%%%%%%%%%%%%%%%%%%%%%%%%%%%%%%%%%%%%%%%%%%%%%%%%%%%%%%%%%%%%%%%%%%%%%%%%%%%%%%%%%%%

%%%%%%%%%%%%%%%%%%%%%%%%%%%%%%%%%%%%%%%%%%%%%%%%%%%%%%%%%%%%%%%%%%%%%%%%%%%%%%%%%%%%%%%%%%%%%%%%%%% 
%%%%%%%%%%%%%%%%%%%%%%%%%%%%%%%%%%%%%%%%%%%%%%%%%%%%%%%%%%%%%%%%%%%%%%%%%%%%%%%%%%%%%%%%%%%%%%%%%%%

%%%%%%%%%%%%%%%%%%%%%%%%%%%%%%%%%%%%%%%%%%%%%%%%%%%%%%%%%%%%%%%%%%%%%%%%%%%%%%%%%%%%%%%%%%%%%%%%%%%
%%%%%%%%%%%%%%%%%%%%% VELOCITY AUTOCORRELATIONS FIGURE, low density   %%%%%%%%%%%%%%%%%%%%%%%%%%%%%
%%%%%%%%%%%%%%%%%%%%%%%%%%%%%%%%%%%%%%%%%%%%%%%%%%%%%%%%%%%%%%%%%%%%%%%%%%%%%%%%%%%%%%%%%%%%%%%%%%%

%%%%%%%%%%%%%%%%%%%%%%%%%%%%%%%%%%%%%%%%%%%%%%%%%%%%%%%%%%%%%%%%%%%%%%%%%%%%%%%%%%%%%%%%%%%%%%%%%%% 
%%%%%%%%%%%%%%%%%%%%%%%%%%%%%%%%%%%%%%%%%%%%%%%%%%%%%%%%%%%%%%%%%%%%%%%%%%%%%%%%%%%%%%%%%%%%%%%%%%%

\section{Discussion}
\label{sec:discussion}

We have studied the nearly-2D dynamics of a system of rolling (inelastic) spheres. The dynamics of
the set of spheres is activated by means of the turbulent vortexes that originate out of an air
upflow past the spheres. As we have seen, the phase behavior of the system is very complex, and we
have been able to detect an arrest phase (particles that remain still or static for low energy
input), a glass phase (disordered lattice of Brownian particles with sporadic jumps to other lattice
positions), a liquid (completely disordered phase) and a hexagonal crystal. In particular, the glass
phase appears at very low densities, which to our knowledge is a very rare situation
\cite{Rodriguez-Rivas2019}. This undoubtedly is due to long-ranged repulsive forces between the
air-fluidized particles that have been reported previously \cite{OAD05}. However, this kind of low
density granular glass had not been reported previously, to the best of our knowledge.  Moreover,
the glass and the hexagonal crystal can coexist with the liquid (again, glass-liquid coexistence had
not been reported previously in the context of granular dynamics, to the best of our
knowledge). Additionally, the glass can also coexist with the arrest phase, at low air current
(dynamics in the process of activation, again not detected previously). In fact, the dynamics of the
system is so complex that we have been able to detect important qualitative differences in the
behavior of a single phase. For instance, as we mentioned above, the velocities of particles in the
liquid phase can be either strongly anticorrelated at early times or not anticorrelated at all,
depending on the configuration of the system. As another example, the crystal can display vanishing
diffusive exponent (and in any case, the crystal is always strongly subdiffusive, see
Figure~\ref{fig:alpha}).

In general, the diffusive properties of the observed phases are rather different. Both the glass and
the crystal are always subdiffusive, and present anticorrelated velocities. By contrast, the liquid
presents always nearly normal diffusion (departures from normal diffusion can be attributed here
either to measurement error or to limitations of our set-up). Measurement of the granular
temperature field confirms that in all cases of phase coexistence there is a significant energy
non-equipartition. This enhances the idea that these transitions are occurring in states very far
from equilibrium. It may be that because of this (that our experimental configurations seem to be
very far away from equilibrium) that the phase transition scenario here described completely differs
from the KTHNY scenario described for 2D equilibrium systems \cite{S88} and also, under certain
conditions, in non-equilibrium systems such as a vibrated granular monolayer \cite{OU05,KT15} and 2D
active brownian disks \cite{DLSCGP18}. Here, however, this KTHNY scenario is as we said absent and
the hexatic phase has not been observed in any situation, which is a very peculiar situation in the
context of two-dimensional matter. Furthermore, all transitions we observed in this work occur
through phase coexistence, contrary to the scenario of liquid-hexatic-crystal continuous phase
transition without coexistence described in the KTHNY theory. Specifically, the fact that the
hexagonal crystal in our set-up melts by shrinking in size, giving rise a growing liquid and the
transition is not defect-mediated, guarantees the complete absence of a hexatic phase during the
melting process. Therefore, our results differ fundamentally in this aspect from previous results in
both equilibrium and non-equilibrium systems (where the hexatic phase has always been observed, at
least, under certain conditions). It remains for future work to study in more detail the structure
of this intriguing phase behavior.

\authorcontributions{J.F. G.-S. and M. A. L.-C. performed all the experiments. M. A. L.-C. and
  F. V. R. developed the particle tracking and processing codes. M.A. L-.C. prepared all
  figures. All authors participated in the formal analysis of the experimental data. A. R.-R. and
  F. V. R. reviewed and edited the manuscript. Design of the experiment, conceptualization, original
  draft preparation, and supervision was performed by F. V. R. All authors have read and agreed to
  the published version of the manuscript.}

\funding{ We acknowledge funding from the Government of Spain through Agencia Estatal de
  Investigaci\'on (AEI) project No. PID2020-116567GB-C22). A.R.-R. also acknowledges financial
  support from Consejer\'ia de Transformaci\'on Econ\'omica, Industria, Conocimiento y Universidades
  de la Junta de Andaluc\'ia/FEDER for funding through project P20-00816 and FSE through
  post-doctoral grant no. DC00316 (PAIDI 20201).  F. V. R. is supported by the Junta de Extremadura
  grant No. GR21091, partially funded by the ERDF. The APC was funded by the MDPI editorial.}

\dataavailability{Experimental data tables and trajectories are available in a public repository, at:\\
  \texttt{https://doi.org/10.5281/zenodo.7097642}.}

\acknowledgments{The authors are indebted to the Taller de Mec\'anica de la Escuela de Ingenier\'ias
  Industriales for construction contributions to the design of the air table. The authors are also
  thankful to Prof. S. B. Yuste and E. Abad for their important contributions to error measurement
  in the early stages of this project.}

\conflictsofinterest{The authors declare no conflict of interest.} 

 \begin{adjustwidth}{-\extralength}{0cm}
% %\printendnotes[custom] % Un-comment to print a list of endnotes

 \reftitle{References}


\bibliography{ppp2}

\begin{thebibliography}{999}

\bibitem[Jaeger \em{et~al.}(1996)Jaeger, Nagel, and Behringer]{JNB96}
Jaeger, H.M.; Nagel, S.; Behringer, R.
\newblock The physics of granular materials.
\newblock {\em Physics Today} {\bf 1996}, {\em 49},~32.

\bibitem[{de Gennes}(1999)]{dG99}
{de Gennes}, P.G.
\newblock Granular matter: a tentative view.
\newblock {\em Rev. Mod. Phys.} {\bf 1999}, {\em 71},~S374--S382.

\bibitem[Aranson and Tsimring(2006)]{AT06}
Aranson, I.S.; Tsimring, L.S.
\newblock Patterns and collective behavior in granular media: Theoretical
  concepts.
\newblock {\em Rev. Mod. Phys.} {\bf 2006}, {\em 78},~641--692.
\newblock {\url{https://doi.org/10.1103/RevModPhys.78.641}}.

\bibitem[Olafsen and Urbach(1998)]{OU98}
Olafsen, J.S.; Urbach, J.S.
\newblock Clustering, order and collapse in a driven granular monolayer.
\newblock {\em Phys. Rev. Lett} {\bf 1998}, {\em 81},~4369--4372.

\bibitem[Goldhirsch(2003)]{G03}
Goldhirsch, I.
\newblock Rapid Granular Flows.
\newblock {\em Annu. Rev. Fluid Mech.} {\bf 2003}, {\em 35},~267--293.

\bibitem[{Vega Reyes} and Urbach(2009)]{VU09}
{Vega Reyes}, F.; Urbach, J.S.
\newblock Steady base states for Navier-Stokes granular hydrolodynamics with
  boundary heating and shear.
\newblock {\em J. Fluid Mech.} {\bf 2009}, {\em 636},~279.

\bibitem[Gantzounis \em{et~al.}(2014)Gantzounis, Yang, Kevrekidis, and
  Daraio]{LAYKA14}
Gantzounis, G.; Yang, J.; Kevrekidis, P.G.; Daraio, C.
\newblock Granular acoustic switches and logic elements.
\newblock {\em Nat. Commun.} {\bf 2014}, {\em 5},~5311.
\newblock {\url{https://doi.org/10.1038/ncomms6311}}.

\bibitem[Gonz{\'a}lez-Saavedra \em{et~al.}(2020)Gonz{\'a}lez-Saavedra,
  Rodr{\'i}guez-Rivas, L{\'o}pez-Casta{\~{n}}o, and {Vega Reyes}]{GRLV2020}
Gonz{\'a}lez-Saavedra, J.F.; Rodr{\'i}guez-Rivas, {\'A}.;
  L{\'o}pez-Casta{\~{n}}o, M.A.; {Vega Reyes}, F.
\newblock Acoustic Resonances in a Confined Set of Disks.
\newblock In Proceedings of the Traffic and Granular Flow 2019; Zuriguel, I.;
  Garcimartin, A.; Cruz, R., Eds.; Springer International Publishing: Cham,
  2020; pp. 349--355.

\bibitem[Mujica and Soto(2016)]{Mujica2016}
Mujica, N.; Soto, R.
\newblock {Dynamics of noncohesive confined granular media}.
\newblock {\em Environmental Science and Engineering (Subseries: Environmental
  Science)} {\bf 2016}, pp. 445--463.

\bibitem[Zik and Stavans(1991)]{ZS91}
Zik, O.; Stavans, J.
\newblock Self-Diffusion in Granular Flows.
\newblock {\em Europhysics Letters (EPL)} {\bf 1991}, {\em 16},~255–258.
\newblock {\url{https://doi.org/10.1209/0295-5075/16/3/006}}.

\bibitem[Oger \em{et~al.}(1996)Oger, Annic, Bideau, Dai, and Savage]{Oger1996}
Oger, L.; Annic, C.; Bideau, D.; Dai, R.; Savage, S.B.
\newblock {Diffusion of two-dimensional particles on an air table}.
\newblock {\em J. Stat. Phys.} {\bf 1996}, {\em 82},~1047.

\bibitem[Ojha \em{et~al.}(2004)Ojha, Lemieux, Dixon, Liu, and Durian]{OLDLD04}
Ojha, R.P.; Lemieux, P.A.; Dixon, P.K.; Liu, A.J.; Durian, D.J.
\newblock Statistical mechanics of a gas-fluidized particle.
\newblock {\em Nature} {\bf 2004}, {\em 427},~521.

\bibitem[Rosato \em{et~al.}(1987)Rosato, Strandburg, Prinz, and
  Swendsen]{RSPS87}
Rosato, A.; Strandburg, K.J.; Prinz, F.; Swendsen, R.H.
\newblock Why the Brazil nuts are on top: Size segregation of particulate
  matter by shaking.
\newblock {\em Phys. Rev. Lett.} {\bf 1987}, {\em 58},~1038--1040.
\newblock {\url{https://doi.org/10.1103/PhysRevLett.58.1038}}.

\bibitem[Kondic \em{et~al.}(2003)Kondic, Hartley, Tennakoon, Painter, and
  Behringer]{KHTPB03}
Kondic, L.; Hartley, R.R.; Tennakoon, S.G.K.; Painter, B.; Behringer, R.P.
\newblock Segregation by friction.
\newblock {\em EPL} {\bf 2003}, {\em 61},~742.

\bibitem[Jenkins and Yoon(2002)]{JY02}
Jenkins, J.T.; Yoon, D.K.
\newblock Segregation in Binary Mixtures under Gravity.
\newblock {\em Phys. Rev. Lett.} {\bf 2002}, {\em 88},~194301.
\newblock {\url{https://doi.org/10.1103/PhysRevLett.88.194301}}.

\bibitem[Hill \em{et~al.}(1999)Hill, Khakhar, Gilchrist, McCarthy, and
  Ottino]{HKGMO99}
Hill, K.M.; Khakhar, D.V.; Gilchrist, J.F.; McCarthy, J.J.; Ottino, J.M.
\newblock Segregation-driven organization in chaotic granular flows.
\newblock {\em Proceedings of the National Academy of Sciences} {\bf 1999},
  {\em 96},~11701--11706,
  \href{http://xxx.lanl.gov/abs/https://www.pnas.org/doi/pdf/10.1073/pnas.96.21.11701}{{\normalfont
  [https://www.pnas.org/doi/pdf/10.1073/pnas.96.21.11701]}}.
\newblock {\url{https://doi.org/10.1073/pnas.96.21.11701}}.

\bibitem[Melby \em{et~al.}(2005)Melby, {Vega Reyes}, Prevost, Robertson, Kumar,
  Egolf, and Urbach]{Melby2005}
Melby, P.; {Vega Reyes}, F.; Prevost, A.; Robertson, R.; Kumar, P.; Egolf,
  D.A.; Urbach, J.S.
\newblock {The dynamics of thin vibrated granular layers}.
\newblock {\em J. Phys.: Condens. Matter} {\bf 2005}, {\em 17},~S2369.

\bibitem[Eshuis \em{et~al.}(2007)Eshuis, {van der Weele}, {van der Meer}, Bos,
  and Lohse]{EWMBL07}
Eshuis, P.; {van der Weele}, K.; {van der Meer}, D.; Bos, R.; Lohse, D.
\newblock Phase diagram of vertically shaken granular matter.
\newblock {\em Phys. Fluids} {\bf 2007}, {\em 19},~123301.

\bibitem[McLaren \em{et~al.}(2019)McLaren, Kovar, Penn, Müller, and
  Boyce]{MKPM19}
McLaren, C.P.; Kovar, T.M.; Penn, A.; Müller, C.R.; Boyce, C.M.
\newblock Gravitational instabilities in binary granular materials.
\newblock {\em Proceedings of the National Academy of Sciences} {\bf 2019},
  {\em 116},~9263–9268.
\newblock {\url{https://doi.org/10.1073/pnas.1820820116}}.

\bibitem[He \em{et~al.}(2002)He, Meerson, and Doolen]{XMD02}
He, X.; Meerson, B.; Doolen, G.
\newblock Hydrodynamics of thermal granular convection.
\newblock {\em Phys. Rev. E} {\bf 2002}, {\em 65},~030301.
\newblock {\url{https://doi.org/10.1103/PhysRevE.65.030301}}.

\bibitem[Pontuale \em{et~al.}(2017)Pontuale, Gnoli, Puglisi, and {Vega
  Reyes}]{PGPV17}
Pontuale, G.; Gnoli, A.; Puglisi, A.; {Vega Reyes}, F.
\newblock Thermal Convection in Granular Gases with Dissipative Lateral Walls.
\newblock {\em Phys. Rev. Lett.} {\bf 2017}, {\em 117},~098006.

\bibitem[Isobe(2012)]{I12a}
Isobe, M.
\newblock Statistical law of turbulence in granular gas.
\newblock {\em J. Phys. Conf. Ser.} {\bf 2012}, {\em 402},~012041.

\bibitem[Isobe(2003)]{I03}
Isobe, M.
\newblock Velocity statistics in two-dimensional granular turbulence.
\newblock {\em Phys. Rev. E} {\bf 2003}, {\em 68},~040301(R).

\bibitem[Liu and Nagel(1998)]{LN98}
Liu, A.; Nagel, S.
\newblock Jamming is not just cool anymore.
\newblock {\em Nature} {\bf 1998}, {\em 396},~21--2.

\bibitem[Daniels \em{et~al.}(2012)Daniels, Haxton, Xu, Liu, and
  Durian]{Daniels2012}
Daniels, L.J.; Haxton, T.K.; Xu, N.; Liu, A.J.; Durian, D.J.
\newblock {Temperature-pressure scaling for air-fluidized grains near jamming}.
\newblock {\em Physical Review Letters} {\bf 2012}, {\em 108},~1--5,
  \href{http://xxx.lanl.gov/abs/arXiv:1110.5611v1}{{\normalfont
  [arXiv:1110.5611v1]}}.
\newblock {\url{https://doi.org/10.1103/PhysRevLett.108.138001}}.

\bibitem[Lasanta \em{et~al.}(2017)Lasanta, {Vega Reyes}, Prados, and
  Santos]{Lasanta2017}
Lasanta, A.; {Vega Reyes}, F.; Prados, A.; Santos, A.
\newblock {When the Hotter Cools More Quickly: Mpemba Effect in Granular
  Fluids}.
\newblock {\em Phys. Rev. Lett.} {\bf 2017}, {\em 119},~1--6,
  \href{http://xxx.lanl.gov/abs/1611.04948}{{\normalfont [1611.04948]}}.
\newblock {\url{https://doi.org/10.1103/PhysRevLett.119.148001}}.

\bibitem[Keim \em{et~al.}(2019)Keim, Paulsen, Zeravcic, Sastry, and
  Nagel]{KPZSN19}
Keim, N.C.; Paulsen, J.D.; Zeravcic, Z.; Sastry, S.; Nagel, S.R.
\newblock Memory formation in matter.
\newblock {\em Rev. Mod. Phys.} {\bf 2019}, {\em 91},~035002.
\newblock {\url{https://doi.org/10.1103/RevModPhys.91.035002}}.

\bibitem[Prevost \em{et~al.}(2004)Prevost, Melby, Egolf, and Urbach]{PMEU04}
Prevost, A.; Melby, P.; Egolf, D.A.; Urbach, J.S.
\newblock Nonequilibrium two-phase coexistence in a confined granular layer.
\newblock {\em Phys. Rev. E} {\bf 2004}, {\em 70},~050301.
\newblock {\url{https://doi.org/10.1103/PhysRevE.70.050301}}.

\bibitem[Reis \em{et~al.}(2006)Reis, Ingale, and Shattuck]{RIS06}
Reis, P.M.; Ingale, R.A.; Shattuck, M.
\newblock Crystallization of a quasi-two-dimensional granular fluid.
\newblock {\em Phys. Rev. Lett.} {\bf 2006}, {\em 96},~258001.

\bibitem[{Vega Reyes} and Urbach(2008)]{VU08}
{Vega Reyes}, F.; Urbach, J.S.
\newblock Effect of inelasticity on the phase transitions of a thin vibrated
  granular layer.
\newblock {\em Phys. Rev. E} {\bf 2008}, {\em 78},~051301.
\newblock {\url{https://doi.org/10.1103/PhysRevE.78.051301}}.

\bibitem[Castillo \em{et~al.}(2012)Castillo, Mujica, and Soto]{CMS12}
Castillo, G.; Mujica, N.; Soto, R.
\newblock Criticality of a Granular Solid-Liquid-Like Phase Transi- tion.
\newblock {\em Phys. Rev. Lett.} {\bf 2012}, {\em 109},~095701.

\bibitem[N\'eel \em{et~al.}(2014)N\'eel, Rondini, Turzillo, Mujica, and
  Soto]{NRTMS14}
N\'eel, B.; Rondini, I.; Turzillo, A.; Mujica, N.; Soto, R.
\newblock Dynamics of a first-order transition to an absorbing state.
\newblock {\em Phys. Rev. E} {\bf 2014}, {\em 89},~042206.
\newblock {\url{https://doi.org/10.1103/PhysRevE.89.042206}}.

\bibitem[Castillo \em{et~al.}(2015)Castillo, Mujica, and Soto]{CMS15}
Castillo, G.; Mujica, N.; Soto, R.
\newblock Universality and criticality of a second-order granular
  solid-liquid-like phase transition.
\newblock {\em Phys. Rev. E} {\bf 2015}, {\em 91},~012141.
\newblock {\url{https://doi.org/10.1103/PhysRevE.91.012141}}.

\bibitem[{Vega Reyes} \em{et~al.}(2010){Vega Reyes}, Santos, and
  Garz\'o]{VSG10}
{Vega Reyes}, F.; Santos, A.; Garz\'o, V.
\newblock Non-Newtonian Granular Hydrodynamics. What Do the Inelastic Simple
  Shear Flow and the Elastic Fourier Flow Have in Common?
\newblock {\em Phys. Rev. Lett.} {\bf 2010}, {\em 104},~028001.
\newblock {\url{https://doi.org/10.1103/PhysRevLett.104.028001}}.

\bibitem[Rietz \em{et~al.}(2018)Rietz, Radin, Swinney, and Schr\"oter]{RRSS18}
Rietz, F.; Radin, C.; Swinney, H.L.; Schr\"oter, M.
\newblock Nucleation in Sheared Granular Matter.
\newblock {\em Phys. Rev. Lett.} {\bf 2018}, {\em 120},~055701.

\bibitem[Goldhirsch and Zanetti(1993)]{GZ93}
Goldhirsch, I.; Zanetti, G.
\newblock Clustering instability in dissipative gases.
\newblock {\em Phys. Rev. Lett.} {\bf 1993}, {\em 70},~1619--1622.

\bibitem[Prevost \em{et~al.}(2002)Prevost, Egolf, and Urbach]{PEU02}
Prevost, A.; Egolf, D.A.; Urbach, J.S.
\newblock Forcing and Velocity Correlations in a Vibrated Granular Monolayer.
\newblock {\em Phys. Rev. Lett.} {\bf 2002}, {\em 89},~084301.
\newblock {\url{https://doi.org/10.1103/PhysRevLett.89.084301}}.

\bibitem[Mujica and Soto(2016)]{MS16}
Mujica, N.; Soto, R.
\newblock Dynamics of Noncohesive Confined Granular Media.
\newblock In Proceedings of the Recent Advances in Fluid Dynamics with
  Environmental Applications; Klapp, J.; Sigalotti, L.D.G.; Medina, A.;
  L{\'o}pez, A.; Ruiz-Chavarr{\'i}a, G., Eds.; Springer International
  Publishing: Cham,  2016; pp. 445--463.

\bibitem[Kosterlitz and Thouless(1972)]{Kosterlitz1972}
Kosterlitz, J.M.; Thouless, D.J.
\newblock {Long range order and metastability in two dimensional solids and
  superfluids. (Application of dislocation theory)}.
\newblock {\em J. Phys. C} {\bf 1972}, {\em 5},~L124--L126.

\bibitem[Kosterlitz and Thouless(1973)]{KT73}
Kosterlitz, J.M.; Thouless, D.J.
\newblock Ordering, metastability and phase transitions in two-dimensional
  systems.
\newblock {\em J. Phys. C} {\bf 1973}, {\em 6},~1181.

\bibitem[Nelson and Halperin(1979)]{NH79}
Nelson, D.R.; Halperin, B.I.
\newblock Dislocation mediated melting in two dimensions.
\newblock {\em Phys. Rev. B} {\bf 1979}, {\em 19},~2457.

\bibitem[Young(1979)]{Y79}
Young, A.P.
\newblock Melting and the vector Coulomb gas in two dimensions.
\newblock {\em Phys. Rev. B} {\bf 1979}, {\em 19},~1855.

\bibitem[Strandburg(1988)]{S88}
Strandburg, K.J.
\newblock Two-dimensional melting.
\newblock {\em Rev. Mod. Phys.} {\bf 1988}, {\em 60},~161.

\bibitem[Olafsen and Urbach(2005)]{OU05}
Olafsen, J.S.; Urbach, J.S.
\newblock Two-Dimensional Melting Far from Equilibrium in a Granular Monolayer.
\newblock {\em Phys. Rev. Lett.} {\bf 2005}, {\em 95},~098002.

\bibitem[Komatsu and Tanaka(2015)]{KT15}
Komatsu, Y.; Tanaka, H.
\newblock Roles of Energy Dissipation in a Liquid-Solid Transition of
  Out-of-Equilibrium Systems.
\newblock {\em Phys. Rev. X} {\bf 2015}, {\em 5},~031025.

\bibitem[Foerster \em{et~al.}(1994{\natexlab{a}})Foerster, Louge, Chang, and
  Allia]{FLCA94}
Foerster, S.F.; Louge, M.Y.; Chang, H.; Allia, K.
\newblock Measurements of the collision properties of small spheres.
\newblock {\em Physics of Fluids} {\bf 1994}, {\em 6},~1108--1115,
  \href{http://xxx.lanl.gov/abs/https://doi.org/10.1063/1.868282}{{\normalfont
  [https://doi.org/10.1063/1.868282]}}.
\newblock {\url{https://doi.org/10.1063/1.868282}}.

\bibitem[Foerster \em{et~al.}(1994{\natexlab{b}})Foerster, Louge, Chang, and
  Allia]{FLCA_data94}
Foerster, S.F.; Louge, M.Y.; Chang, H.; Allia, K.
\newblock Measurements of the collision properties of small spheres.
  Experimental data table.
\newblock Technical report,  1994.

\bibitem[Olafsen and Urbach(1999)]{OU99}
Olafsen, J.S.; Urbach, J.S.
\newblock Velocity distributions and density fluctuations in a granular gas.
\newblock {\em Phys. Rev. E} {\bf 1999}, {\em 60},~R2468.

\bibitem[Olafsen and Urbach(2001)]{OU01}
Olafsen, J.S.; Urbach, J.S., Experimental observations of non-equilibrium
  distributions and transitions in a {2D} granular gas; Springer-Verlag:
  Berlin, Germany,  2001; Vol. Granular Gases, {\em Lecture Notes in Physics},
  pp. 410--428.

\bibitem[Brey \em{et~al.}(1998)Brey, Dufty, Kim, and Santos]{Brey1998}
Brey, J.J.; Dufty, J.W.; Kim, C.S.; Santos, A.
\newblock {Hydrodynamics for granular flow at low density}.
\newblock {\em Physical Review E} {\bf 1998}, {\em 58},~4638--4653.
\newblock {\url{https://doi.org/10.1103/PhysRevE.58.4638}}.

\bibitem[Brey and Cubero(2001)]{BC01}
Brey, J.J.; Cubero, D., Hydrodynamic transport coefficients of granular gases;
  Springer-Verlag: Berlin, Germany,  2001; Vol. Granular Gases, {\em Lecture
  Notes in Physics}, pp. 59--78.

\bibitem[Bechinger \em{et~al.}(2016)Bechinger, Di~Leonardo, L\"owen,
  Reichhardt, Volpe, and Volpe]{BLLRVV16}
Bechinger, C.; Di~Leonardo, R.; L\"owen, H.; Reichhardt, C.; Volpe, G.; Volpe,
  G.
\newblock Active particles in complex and crowded environments.
\newblock {\em Rev. Mod. Phys.} {\bf 2016}, {\em 88},~045006.
\newblock {\url{https://doi.org/10.1103/RevModPhys.88.045006}}.

\bibitem[Digregorio \em{et~al.}(2018)Digregorio, Levis, Suma, Cugliandolo,
  Gonnella, and Pagonabarraga]{DLSCGP18}
Digregorio, P.; Levis, D.; Suma, A.; Cugliandolo, L.F.; Gonnella, G.;
  Pagonabarraga, I.
\newblock Full Phase Diagram of Active Brownian Disks: From Melting to
  Motility-Induced Phase Separation.
\newblock {\em Phys. Rev. Lett.} {\bf 2018}, {\em 121},~098003.
\newblock {\url{https://doi.org/10.1103/PhysRevLett.121.098003}}.

\bibitem[Ojha \em{et~al.}(2005)Ojha, Abate, and Durian]{OAD05}
Ojha, R.P.; Abate, A.R.; Durian, D.J.
\newblock Statistical characterization of the forces on spheres in an upflow of
  air.
\newblock {\em Phys. Rev. E} {\bf 2005}, {\em 71},~016313.
\newblock {\url{https://doi.org/10.1103/PhysRevE.71.016313}}.

\bibitem[Batchelor(1974)]{B74}
Batchelor, G.K.
\newblock Transport properties of two-phase materials with random structure.
\newblock {\em Ann. Rev. Fluid Mech.} {\bf 1974}, {\em 6},~227--255.

\bibitem[Ojha \em{et~al.}(2005)Ojha, Abate, and Durian]{Ojha2005}
Ojha, R.P.; Abate, A.R.; Durian, D.J.
\newblock {Statistical characterization of the forces on spheres in an upflow
  of air}.
\newblock {\em Phys. Rev. E} {\bf 2005}, {\em 71},~016313.

\bibitem[L\'opez-Casta\~no \em{et~al.}(2021)L\'opez-Casta\~no,
  Gonz\'alez-Saavedra, Rodr\'{\i}guez-Rivas, Abad, Yuste, and {Vega
  Reyes}]{LGRAYV21}
L\'opez-Casta\~no, M.A.; Gonz\'alez-Saavedra, J.F.; Rodr\'{\i}guez-Rivas, A.;
  Abad, E.; Yuste, S.B.; {Vega Reyes}, F.
\newblock Pseudo-two-dimensional dynamics in a system of macroscopic rolling
  spheres.
\newblock {\em Phys. Rev. E} {\bf 2021}, {\em 103},~042903.
\newblock {\url{https://doi.org/10.1103/PhysRevE.103.042903}}.

\bibitem[Abate and Durian(2005)]{Abate2005}
Abate, A.R.; Durian, D.J.
\newblock {Partition of energy for air-fluidized grains}.
\newblock {\em Phys. Rev. E} {\bf 2005}, {\em 72},~031305.

\bibitem[L{\'o}pez-Casta{\~{n}}o \em{et~al.}(2020)L{\'o}pez-Casta{\~{n}}o,
  Gonz{\'a}lez-Saavedra, Rodr{\'i}guez-Rivas, and {Vega Reyes}]{LGRV20}
L{\'o}pez-Casta{\~{n}}o, M.A.; Gonz{\'a}lez-Saavedra, J.F.;
  Rodr{\'i}guez-Rivas, {\'A}.; {Vega Reyes}, F.
\newblock Statistical Properties of a Granular Gas Fluidized by Turbulent Air
  Wakes.
\newblock In Proceedings of the Traffic and Granular Flow 2019; Zuriguel, I.;
  Garcimartin, A.; Cruz, R., Eds.; Springer International Publishing: Cham,
  2020; pp. 397--403.

\bibitem[Koyama \em{et~al.}(2021)Koyama, Matsuno, and Noguchi]{KMN21}
Koyama, S.; Matsuno, T.; Noguchi, T.
\newblock Anomalous diffusion in a monolayer of lightweight spheres fluidized
  in air flow.
\newblock {\em Phys. Rev. E} {\bf 2021}, {\em 104},~054901.
\newblock {\url{https://doi.org/10.1103/PhysRevE.104.054901}}.

\bibitem[Maw \em{et~al.}(1981)Maw, Barber, and Fawcett]{MBF81}
Maw, N.; Barber, J.R.; Fawcett, J.N.
\newblock {The Role of Elastic Tangential Compliance in Oblique Impact}.
\newblock {\em Journal of Lubrication Technology} {\bf 1981}, {\em
  103},~74--80,
  \href{http://xxx.lanl.gov/abs/https://asmedigitalcollection.asme.org/tribology/article-pdf/103/1/74/5796519/74\_1.pdf}{{\normalfont
  [https://asmedigitalcollection.asme.org/tribology/article-pdf/103/1/74/5796519/74\_1.pdf]}}.
\newblock {\url{https://doi.org/10.1115/1.3251617}}.

\bibitem[Taneda(1978)]{T78}
Taneda, S.
\newblock Visual observations of the flow past a sphere at Reynolds numbers
  between $10^4$ and $10^6$.
\newblock {\em J. Fluid Mech.} {\bf 1978}, {\em 85},~187--192.

\bibitem[{Van Dyke}(1982)]{vD82}
{Van Dyke}, M.
\newblock {\em An album of fluid motion}; The Parabolic Press: Stanford, CA,
  USA,  1982.

\bibitem[Montanero \em{et~al.}(1999)Montanero, Garz\'o, Santos, and
  Brey]{MGSB99}
Montanero, J.M.; Garz\'o, V.; Santos, A.; Brey, J.J.
\newblock Kinetic theory of simple granular shear flows of smooth hard spheres.
\newblock {\em J. Fluid Mech.} {\bf 1999}, {\em 389},~391--411.

\bibitem[{OpenCV}()]{opencv}
{OpenCV}.
\newblock \texttt{https://opencv.org/}.

\bibitem[{Allan, D. \textit{et al.}}(2019)]{dan_allan_2019_3492186}
{Allan, D. \textit{et al.}}.
\newblock soft-matter/trackpy: Trackpy v0.4.2,  2019.

\bibitem[{Vega Reyes} \em{et~al.}(2022){Vega Reyes}, L{\'o}pez-Casta{\~{n}}o,
  and Rodr\'{\i}guez-Rivas]{ppp2_code}
{Vega Reyes}, F.; L{\'o}pez-Casta{\~{n}}o, M.A.; Rodr\'{\i}guez-Rivas, A.,
  2022.

\bibitem[Kanatani(1979)]{K79}
Kanatani, K.I.
\newblock A micropolar continuum theory for the flow of granular materials.
\newblock {\em Int. J. Engng. Sci.} {\bf 1979}, {\em 17},~419--432.

\bibitem[Desmond and Weeks(2009)]{Desmond2009}
Desmond, K.W.; Weeks, E.R.
\newblock {Random close packing of disks and spheres in confined geometries}.
\newblock {\em Physical Review E - Statistical, Nonlinear, and Soft Matter
  Physics} {\bf 2009}, {\em 80},~1--11,
  \href{http://xxx.lanl.gov/abs/0903.0864}{{\normalfont [0903.0864]}}.
\newblock {\url{https://doi.org/10.1103/PhysRevE.80.051305}}.

\bibitem[Rodr{\'{i}}guez-Rivas \em{et~al.}(2019)Rodr{\'{i}}guez-Rivas,
  Romero-Enrique, and Rull]{Rodriguez-Rivas2019}
Rodr{\'{i}}guez-Rivas, A.; Romero-Enrique, J.M.; Rull, L.F.
\newblock {Molecular simulation study of the glass transition in a soft
  primitive model for ionic liquids}.
\newblock {\em Mol. Phys.} {\bf 2019}, {\em 117},~3941--3956.
\newblock {\url{https://doi.org/10.1080/00268976.2019.1674935}}.

\bibitem[Metzler \em{et~al.}(2014)Metzler, Jeon, Cherstvy, and Barkai]{MJCB14}
Metzler, R.; Jeon, J.H.; Cherstvy, A.G.; Barkai, E.
\newblock Anomalous diffusion models and their properties: non-stationarity{,}
  non-ergodicity{,} and ageing at the centenary of single particle tracking.
\newblock {\em Phys. Chem. Chem. Phys.} {\bf 2014}, {\em 16},~24128--24164.
\newblock {\url{https://doi.org/10.1039/C4CP03465A}}.

\bibitem[Kranz \em{et~al.}(2010)Kranz, M.~Sperl, and Zippelius]{KSZ10}
Kranz, W.T.; M.~Sperl, .; Zippelius, A.
\newblock Glass Transition for Driven Granular Fluids.
\newblock {\em Phys. Rev. Lett.} {\bf 2010}, {\em 104}.

\bibitem[L\'opez-Casta{\~{n}}o \em{et~al.}(2022)L\'opez-Casta{\~{n}}o, {Vega
  Reyes}, and Rodr\'iguez-Rivas]{LVR22}
L\'opez-Casta{\~{n}}o, M.A.; {Vega Reyes}, F.; Rodr\'iguez-Rivas, A.
\newblock Diffusive regimes in a two-dimensional chiral fluid.
\newblock {\em Comms. Phys.} {\bf 2022}, {\em to appear}.
\newblock preprint: arXiv:2202.08920.

\end{thebibliography}

\end{adjustwidth}
\end{document}